\documentclass[aps,prd,10pt,nofootinbib,twocolumn,eqsecnum,
               showpacs,showkeys,superscriptaddress,preprintnumbers,
               altaffilletter,floatfix]{revtex4-1}

\usepackage[utf8]{inputenc}
\usepackage[T1]{fontenc}
\usepackage[english]{babel}

\usepackage{amsmath,amssymb,amsfonts,amsbsy}

\usepackage{graphicx}
\usepackage{dcolumn}
\usepackage{multirow}
\usepackage[caption=false]{subfig}  

\usepackage{soul}      
\usepackage{color}
\usepackage{rotating}

\usepackage[colorlinks=true,
            linkcolor=blue,
            citecolor=cyan,
            urlcolor=cyan,
            filecolor=magenta]{hyperref}

\usepackage{microtype}
\DeclareUnicodeCharacter{0327}{\c{}}  

\newcommand{\be}{\begin{equation}}
\newcommand{\ee}{\end{equation}}
\newcommand{\bea}{\begin{eqnarray}}
\newcommand{\eea}{\end{eqnarray}}

\begin{document}

\title{ Addressing the Hubble Tension: Insights from Reversible and Irreversible Thermodynamic Processes}
\author{Hussain Gohar}
\email{hussain.gohar@usz.edu.pl}
\affiliation{Institute of Physics, University of Szczecin, Wielkopolska 15, 70-451 Szczecin, Poland}

\date{\today}

\begin{abstract}
\noindent
We investigate reversible and irreversible thermodynamic processes in cosmology and their impact on the Hubble tension through modifications to early and late-time expansion history. Gravitationally induced adiabatic matter creation/annihilation are treated as irreversible processes, while energy exchange between the cosmic bulk and horizon is modeled as reversible. We propose two thermodynamically interacting scenarios: Model I considers matter creation/annihilation across all species with energy transfer to effective entropic dark energy, while Model II focuses on dark matter creation/annihilation with energy flow from baryonic matter and radiation. Both incorporate the generalized first law of thermodynamics with matter creation/annihilation governed by $\Gamma(t) = \Gamma_0 H$ and energy transfer quantified by parameter $\gamma$.
We perform observational analysis using Pantheon$+$, cosmic microwave background distance priors, baryon acoustic oscillations, gamma-ray bursts, cosmic chronometers, with and without SH$0$ES measurements. When SH$0$ES data are included, matter annihilation scenarios ($\Gamma_0 < 0$) become statistically preferred, yielding $H_0 = 71.75 \pm 0.79$ km s$^{-1}$ Mpc$^{-1}$ (Model I) and $H_0 = 71.06 \pm 0.81$ km s$^{-1}$ Mpc$^{-1}$ (Model II), achieving $1.2\sigma$ and $1.8\sigma$ agreement with the SH$0$ES measurement of $H_0 = 73.17 \pm 0.86$ km s$^{-1}$ Mpc$^{-1}$. Matter creation ($\Gamma_0 > 0$) or pure energy flow ($\Gamma(t) = 0$) scenarios show no such improvement. However, without SH$0$ES data, information criteria show no preference for the thermodynamically interacting models over $\Lambda$CDM, indicating the models' performance depends critically on local distance ladder calibration.
For matter annihilation scenarios with energy flow, effective entropic dark energy exhibits dynamical evolution, mimicking radiation and matter before recombination, then transitioning through quintessence toward the cosmological constant today. These results demonstrate that thermodynamically motivated interactions provide a theoretically consistent framework that accommodates tension alleviation when calibrated with SH$0$ES measurements, while highlighting the broader challenge of reconciling early-universe and local observational constraints.
\end{abstract}

\maketitle
\section{Introduction}
\noindent
The Lambda Cold Dark Matter ($\Lambda$CDM) model provides a robust explanation for the accelerated expansion of the universe and aligns with high-precision observational data \cite{Perlmutter:1998np,Riess:1998cb,Hinshaw:2012aka,Planck:2018vyg,eBOSS:2020yzd,Brout:2022vxf,DES:2022ccp,DESI:2025zgx}. However, it faces several theoretical and observational challenges \cite{Perivolaropoulos:2021jda, Bull:2015stt, Freedman:2021ahq}, most notably its inability to elucidate the origins of dark energy \cite{Li:2011sd,Nojiri:2006ri,Ishak:2018his} and dark matter \cite{Arbey:2021gdg, Cirelli:2024ssz}. From a data-centric viewpoint, the principal issue lies in the prediction of the current value of the Hubble parameter $H_0$ by $\Lambda$CDM using cosmic microwave background (CMB) data \cite{Planck:2018vyg} compared to direct observations of $H_0$ from SH$0$ES \cite{Riess:2020fzl, Riess:2021jrx, Breuval:2024lsv}. The discrepancy first surfaced with Planck satellite data, highlighting a mismatch between the base $\Lambda$CDM estimation of $H_0$ and the evaluation of the Cepheid distance ladder by the SH$0$ES collaboration. This {\it Hubble tension} has intensified as SH$0$ES' recent measurements \cite{Breuval:2024lsv} show $H_0 = 73.17 \pm 0.86 \, \text{km} \, \text{s}^{-1} \, \text{Mpc}^{-1}$, which is $5.8\sigma$ above Planck's base $\Lambda$CDM value of $H_0 = 67.4 \pm 0.5 \, \text{km} \, \text{s}^{-1} \, \text{Mpc}^{-1}$.
A wide variety of solutions have been proposed to address this tension \cite{Vagnozzi:2019ezj, Schoneberg:2021qvd, DiValentino:2020zio, DiValentino:2021izs, Verde:2019ivm, DiValentino:2025sru}. Among them are holographic dark energy models \cite{Dai:2020rfo}, dynamical dark energy models \cite{Yang:2018qmz}, interacting dark energy models \cite{DiValentino:2019ffd, Kumar:2019wfs}, and early dark energy models \cite{Poulin:2018cxd, Kamionkowski:2022pkx}, among others. Reference \cite{DiValentino:2021izs} presents a systematic classification of these solutions according to how well they resolve the $H_0$ tension combining different data sets, taking into account the $1\sigma$, $2\sigma$, and $3\sigma$ confidence levels. 
An approach to addressing the {\it Hubble tension} involves altering the post-recombination expansion history to raise the $H_0$ value, termed as {\it late time solutions}. Contrarily, {\it early time solutions} modify the expansion history before recombination, thereby adjusting $H_0$ to resolve the tension.

In this article, we investigate a new way to address the {\it Hubble tension}
by considering thermodynamic reversible and irreversible processes to the energy momentum tensor. These modifications change the energy-momentum tensor of the perfect fluid by incorporating additional terms coming from the gravitationally induced adiabatic creation/annihilation of matter (irreversible process) and the energy exchange between the bulk and the cosmological horizon (reversible process). As a result, the Friedmann, acceleration, and continuity equations are modified, and these changes are responsible for the present accelerated expansion of the universe. Furthermore, these modifications provide a promising way to alleviate the {\it Hubble tension}.

Thermodynamic processes, encompassing both reversible and irreversible, hold a pivotal role within cosmology. Irreversible processes were initially introduced to the cosmological framework by Prigogine et al. \cite{Prigogine:1989zz, Prigogine1988} and have subsequently been employed to describe both the universe's early and late-time accelerated expansion phases of the universe \cite{Lima:1987fj,Lima:1992np,Calvao:1991wg,Lima:1995xz,Alcaniz:1999hu,Abramo:1996ip,Zimdahl:1999tn,Lima:2014hda,Jesus:2011ek,Pan:2016jli,Pan:2018ibu,Freaza:2002ic,Halder:2025eze,SolaPeracaula:2019kfm, Harko:2015pma,Gohar:2020bod}. These processes entail the adiabatic creation of matter, culminating in the generation of irreversible entropy. The mechanism of energy exchange, construed as a reversible process, between the universe's bulk and its cosmological horizon is intrinsically rooted in the holographic principle, which interrelates entropy, temperature, and energy with the horizon, indicating a reversible modality of energy transfer. Foundational investigations, including \cite{Freese:1986dd,Lima:1995ea,Maia:2001zu,Barrow:2006hia,Szydlowski:2005ph,Szydlowski:2005kv,Lima:1995kd, Calvao:1992wr}, have examined phenomena such as vacuum decay and fluid interactions. This study endeavors to integrate both reversible and irreversible processes to deepen the comprehension of the universe's evolution.

The functions pertaining to the rate of matter creation and the entropy of the horizon, with the respective horizon temperature, are integral to our analysis. Nonetheless, the exact mechanisms underpinning gravitational particle production remain inadequately comprehended within the framework of quantum field theory \cite{Parker:1968mv,Parker:1971pt,Zeldovich:1977vgo}. We employ a specific phenomenological function for matter creation rates; for considerations of other general phenomenological functions, see \cite{Gohar:2020bod, Halder:2025eze}. In relation to the flow of energy, we utilize the conventional Hawking temperature alongside our recently introduced generalized horizon entropy \cite{Gohar:2023lta}. We outline scenarios of matter creation and energy flow across the horizon, thereby formulating distinct cosmological interaction scenarios. These scenarios offer an alternative to the conventional dark energy model and potentially alleviate the Hubble tension. The assessment of these models in light of current cosmological data enables us to impose constraints on parameters.

The structure of the paper is as follows: Thermodynamic processes in relation to cosmological fluids are covered in Section II. In Section III, various cosmological scenarios are introduced to examine the Hubble tension, along with the modified Friedmann, acceleration and continuity equations based on Section II. Section IV examines the data and the statistical technique used for data analysis. Section V presents the findings and Section VI provides the final conclusion and a summary of the article.
\section{Reversible and Irreversible Processes}
\noindent
We consider a spatially flat, homogeneous, and isotropic universe that functions as an open thermodynamic system characterized by a specific volume $V=V(t)$ and a defined particle number $N=N(t)$. Within this framework, we posit the occurrence of matter creation in the universe coupled with an energy exchange between the bulk and its boundary. Under these conditions, the first law of thermodynamics can be expressed as
\be
\frac{d}{dt}(\epsilon V) + \left(p+p_c\right)\frac{dV}{dt} = \frac{dQ}{dt}, \label{Ceq1} 
\ee
where the internal energy is denoted by $E=\epsilon V$, the energy density is represented by $\epsilon=\rho c^2$ that incorporates the speed of light $c$, and the thermodynamic pressure is indicated by $p$. The creation pressure for adiabatic matter creation is introduced in \cite{Prigogine:1989zz, Prigogine1988} as 
\be \label{P_c}
p_c=-\frac{(\epsilon + p)V\Gamma(t)}{\dot V}.
\ee
Here, the rate of matter creation is specified by $\Gamma(t)$, which is associated with $\dot n+3\frac{\dot a}{a}n=n\Gamma(t)$, along with the number density of particles $n=N/V$. $\Gamma(t)>0$ corresponds to matter creation, while $\Gamma(t)<0$ corresponds to annihilation. On the right-hand side of equation (\ref{Ceq1}), the heat flow $dQ_{rev}=TdS_{rev}$ across the horizon is delineated, leading to a total entropy change $dS= dS_{rev}+dS_{irr}$ with $dS_{irr}\ge 0$, where $S_{rev}$ and $S_{irr}$ correspond to reversible and irreversible entropies. Furthermore, we assume an isothermal quasistatic process for energy exchange between the bulk and the horizon; therefore, the temperature at the horizon, matches the bulk temperature. We use $dQ_{rev}=T_{h}dS_m$, with an approximate Hawking temperature $T_h =\hbar c/ (2 \pi k_B L)$, where $\hbar$ represents the reduced Planck constant and we consider the boundary as the Hubble horizon $L = c/H$ with the Hubble function $H$. In addition to this, we use our recently introduced generalized horizon entropy \cite{Gohar:2023lta} $S_m$ based on generalized mass to horizon relation, defined as
\be S_m = \frac{2\pi k_B m \gamma}{(m+1)l_p^2}L^{m+1}, \label{S_m}
\ee with $m$ as a nonnegative real number, $\gamma$ a free positve parameter with dimensions of $[length]^{1-m}$, $l_p$ as Planck length, and $k_B$ as the Boltzmann constant. Using $V(t)=\frac{4}{3}\pi L^3$, $S_m$, and $T_h$, the first law (\ref{Ceq1}) is reformulated as
\be \label{Ceq2}
\dot\rho+3H\left(\rho+\frac{p_{eff}}{c^2}\right)= -\left(\frac{3\gamma mc^{m-1}}{4\pi G}\right)H^{2-m}\dot H,
\ee where effective pressure $p_{eff} = p + p_c$ is introduced with the definition of creation pressure $p_c$.
Indeed, this equation represents a generalization of the continuity equations examined in the framework of matter creation cosmologies \cite{Lima:2015xpa,Lima_2010}, through the incorporation of energy transfer from the bulk to the horizon. The role of $\gamma$ is to quantify this energy flow. Moreover, the aforementioned equation pertains to a single fluid model with the creation of that specific fluid. However, scenarios involving multiple fluids with distinct matter creation rates for each fluid or for one of them exclusively can also be envisaged. In the subsequent section, we will explore scenarios involving multiple fluids, considering matter creation for all fluids or for only a single fluid, alongside the energy exchange among fluid species as quantified by $\gamma$. 

We wish to address the topic of generalized horizon entropy $S_m$. For comprehensive details and justifications, see our recent articles \cite{Gohar:2023lta, Gohar:2023hnb, Cimdiker:2022ics}. Traditionally, the association between entropy and temperature on horizons has been established through Bekenstein entropy, which is proportional to the horizon area and is characterized by the Hawking temperature, defined in terms of surface gravity. Various generalizations have been proposed, such as Tsallis-Cirto \cite{Tsallis:2012js} (based on Tsallis nonextensive statistics), Barrow \cite{Barrow:2020tzx} (inspired by fractal horizons), and R\'enyi \cite{Alonso-Serrano:2020hpb} (motivated from quantum information theory). These, however, reveal inconsistencies with the Hawking temperature when examined under the holographic principle \cite{Gohar:2023hnb,Cimdiker:2022ics}. To address this issue, we have introduced a novel generalized entropy $S_m$ \cite{Gohar:2023lta}, derived from a generalized mass-to-horizon relation\footnote{Extensions of this relation that incorporate quantum gravitational effects and entanglement-entropy motivated corrections have been proposed in \cite{Gohar:2025yfx}.}
\be
M=\gamma\frac{c^2}{G}L^m \label{ML}
\ee
and the Hawking temperature $T_h$, employing the Clausius relation. This entropy reduces to the Bekenstein entropy under certain conditions $\gamma = m =1$, and $m$ can also be rescaled to align with other nonstandard entropies, such as Barrow and Tsallis-Cirto. Notably, $S_m$ maintains thermodynamic consistency with $T_h$ when utilized in conjunction with the generalized mass-to-horizon relation. The generalized mass to horizon relation plays very important role in order to have thermodynamic consistency, therefore one should modify this relation to use Hawking temperature with any other nonstandard entropy. For example, for Tsallis-Cirto and Barrow entropies, one must rescale $m$ with Tsallis-Cirto and Barrow parameters as suggested in \cite{Gohar:2023lta,Gohar:2025yfx}.
\section{Modified Friedmann Equations}
\noindent
We consider a spatially flat Friedmann–Lemaître–Robertson–Walker (FLRW) metric $g_{\mu \nu}$ and present a modified energy-momentum tensor $\tilde{T}_{\mu \nu}$ of the following form $\tilde{T}_{\mu \nu} = T_{\mu \nu} - \frac{\Lambda_{\text{rev}}(t)}{8\pi G} g_{\mu\nu}$. This tensor accounts for changes brought about by matter creation and energy transfer across the cosmological horizon.  
$T_{\mu \nu}$ comprises the standard isotropic pressure, \( p \), of a perfect fluid as well as an extra creation pressure, \( p_c \), associated to matter creation. Moreover, a time-dependent cosmological term, \( \Lambda_{\text{rev}}(t) \), represents the influence of reversible energy flow and is driven by the energy exchange between the bulk and the horizon.  The divergence of the modified energy-momentum tensor, \( \tilde{T}_{\mu \nu} \), inevitably leads to a continuity equation with the definition of \( \Lambda_{\text{rev}}(t) \) coming from the right side of equation (\ref{Ceq2}).
 The modified energy-momentum tensor $\tilde{T}_{\mu \nu}$ thus describes a thermodynamic open system that is expanding nonadiabatically, including the creation of matter and the transfer of energy from the bulk to the horizon.  For a multi-fluid scenario, we obtain the modified Friedmann and acceleration equations by solving the Einstein field equations as follows:
\begin{align} 
&H^2 =\frac{8\pi G}{3}\sum_{i}\rho_i +\frac{8\pi G}{3}\rho_e\label{eq:Fa11_no_f_t}\\
&\frac{\ddot a}{a}=-\frac{4\pi G}{3}\sum_{i}\left(\rho_i+\frac{3p_{eff,_i}}{c^2}\right) + \frac{8\pi G}{3}\rho_e \, , \label{eq:Aa11_no_g_t} \\
&\sum_{i}\left[\dot \rho_i+3H\left(\rho_i+\frac{p_{eff,_i}}{c^2}\right)\right] = -\dot\rho_e \, ,\label{eq:Ca11_no_f_t_g_t}
\end{align}
where the index $i$ runs for the total matter and radiation and we introduce the effective entropic dark energy density 
\be \label{rhoe}
\rho_e = \frac{3}{8\pi G}\left(\frac{2 m \gamma}{3-m}c^{m-1}H^{3-m}\right)+c_0,
\ee
where $c_0$ is an integration constant. It is noteworthy that the aforementioned set of equations effectively integrates the matter creation cosmologies \cite{Lima:2015xpa,Lima_2010} with the energy flow cosmologies \cite{Barrow:2006hia}, as governed by the matter creation function $\Gamma(t)$ and the energy flow parameter $\gamma$. In the case of $\Gamma(t)=\gamma =0$, these equations simplify to those represented by the standard $\Lambda$CDM model. It should be emphasized that these equations possess a high degree of generality concerning matter creation scenarios and energy flux. For instance, one might consider gravitationally induced adiabatic dark matter creation while all other species adhere to the standard continuity equations. Additionally, one may contemplate the adiabatic creation of all species of the universe with different creation rate functions. Intriguingly, scenarios involving solely matter creation in the absence of energy flow, as addressed by $\Gamma(t)>0$ for $\rho +p>0$, elicit a negative creation pressure, which might account for the current accelerated expansion of the universe. However, the data will tell us about the sign of $\Gamma(t)$.

We shall subsequently employ $m=1$ in conjunction with a phenomenological function corresponding to $\Gamma(t)= \Gamma_0 H$. Interestingly, this choice also encompasses the running vacuum models investigated in \cite{ Sola:2013gha, SolaPeracaula:2022hpd, Moreno-Pulido:2020anb, Moreno-Pulido:2022phq, SolaPeracaula:2023swx, MorenoPulido:2023jsc, SolaPeracaula:2017esw, Sola:2016ecz, Sola:2017jbl, SolaPeracaula:2016qlq}. Our aim is to preserve the definition of the standard Bekenstein entropy, i.e. for $m=1$. For nonstandard entropies as discussed in the preceding section, such as the Tsallis-Cirto entropy \cite{Tsallis:2012js} and the Barrow entropy \cite{Barrow:2020tzx}, a variety of parameterizations may be adopted for $m$ \cite{Gohar:2023lta, Gohar:2025yfx}. Furthermore, alternative functional forms may be ascribed to $\Gamma(t)$\cite{Gohar:2020bod,Halder:2025eze}.The aforementioned selection of $\Gamma(t)$ has been employed in the literature due to its simplicity, however, extending beyond this function is not within the scope of this work. 


In the subsequent subsections, we examine two distinct scenarios: Firstly, those instances wherein matter and radiation are impacted by the creation rate function, with energy being transferred from radiation and matter to the effective entropic dark energy. Secondly, we focus solely on the scenario involving the creation of cold dark matter exclusively. In this context, it is presumed that baryons and radiation adhere to the standard continuity equation, with energy transferring from baryons and radiation to dark matter and, subsequently, energy flowing from all three constituents to the effective entropic dark energy.
\subsection*{Model - I}
\noindent
For the first consideration, the continuity equations for each constituent can be expressed as 
\begin{align} 
&\dot \rho_m+3H\left(\rho_m+\frac{p_{eff,m}}{c^2}\right)\ = 3\gamma H(1-\frac{\Gamma_0}{3})\rho_m, \label{rho_m}\\
&\dot \rho_r+3H\left(\rho_r+\frac{p_{eff,r}}{c^2}\right)\ = 4\gamma H(1-\frac{\Gamma_0}{3})\rho_r, \label{rho_R} \\
&\dot \rho _e = -3H \gamma(1-\frac{\Gamma_0}{3})\left(\rho_m +\frac{4}{3} \rho_r \right), \label{rho_e1}
\end{align}
where $p_{eff,i}= p_i+p_{c,i}$ with the creation pressure $p_{c_i}$ for total matter and radiation, defined in equation (\ref{P_c}) for the matter creation function $\Gamma(t)=\Gamma_0H$.  Here, we have 
$\dot H=-4\pi G(1-\frac{\Gamma_0}{3})\left(\rho_m +\frac{4}{3} \rho_r \right)$ in (\ref{eq:Ca11_no_f_t_g_t}) to get above equations. 
We have utilized equations (\ref{eq:Fa11_no_f_t}) and (\ref{eq:Aa11_no_g_t}) to derive $\dot H$. Subsequently, Equation (\ref{rhoe}) for $m=1$ was substituted into equation (\ref{eq:Ca11_no_f_t_g_t}), and the continuity equations for matter and radiation were separated by incorporating the right-hand side of equation (\ref{eq:Ca11_no_f_t_g_t}).
In this scenario, the total matter density $\rho_m=\rho_{dm}+\rho_b$ is composed of the densities of dark matter $\rho_{dm}$ and baryonic matter $\rho_b$. As indicated by the equations above, matter and radiation do not engage in direct interaction; instead, their interaction is mediated indirectly through effective entropic dark energy $\rho_e$. Nonetheless, the effective entropic dark energy interacts with both matter and radiation through the transfer of energy from these components.

We rewrite the above continuity equations (\ref{rho_m}) and (\ref{rho_R}), using $p_{eff,i}$ for matter and radiation with $\Gamma(t)=\Gamma_0H$, as 
\be
\dot\rho_i+3H\left(1+w_{eff,i}\right)\rho_i = 0,
\ee 
where we have introduced an effective equation of state $w_{eff,i}$, given by
$w_{eff,i} = \beta_i -1,~~ \text{with}~~ \beta_i =\alpha(1+w_i)$, and the parameter $\alpha$ is defined as
\be \label{alpha}
\alpha=(1-\gamma)(1-\frac{\Gamma_0}{3}).
\ee
Solving the above continuity equations for matter and radiation, we get
$
\rho_{m} = \rho_{m,0} a^{-3 \alpha}, ~~ \rho_{r}=\rho_{r,0}a^{-4 \alpha} \label{rho_mr}
.$ 
and the solution for $\rho_e$ can be easily found as 
\begin{align}
\rho_e &= \frac{\gamma}{1-\gamma} \bigg[ 
\rho_{m,0} \left(a^{-3 \alpha}-1\right)
+ \rho_{r,0} \left(a^{-4 \alpha}-1\right) \bigg] + \rho_{e,0}, \label{rho_e}
\end{align}
where $\rho_{m,0}$, $\rho_{r,0}$ and $\rho_{e,0}$
are current values at $a=1$.
Employing equation (\ref{rho_e1}), we can write the equation of state $w_e$ in terms of pressure $p_e = w_{e}\rho_e $ for the effective entropic dark energy $rho_e$.
\be
w_{e}= - 1 + \gamma(1-\frac{\Gamma_0}{3})\frac{H_0^2}{H^2}\frac{\Omega_{m,0}a^{-3\alpha}+\frac{4}{3}\Omega_{r,0}a^{-4\alpha}}{\Omega_{e}(a)},
\ee
where we have introduced the dimensionless density parameters $\Omega_i(a)=\frac{8\pi G}{3H^2(a)}\rho_{i}(a)$ with the present value defined at $a=1$, i.e., $\Omega_{i,0}=\frac{8\pi G}{3H_0^2}\rho_{i,0}$. Now, the modified continuity equation for $\rho_e$ becomes
\be
\dot\rho_e+3H\left(1+w_{e}\right)\rho_e = 0.
\ee
Then, employing $\rho_m$, $\rho_r$ and $\rho_e$ in equation (\ref{eq:Fa11_no_f_t}), the normalized Friedmann equation reads as
\begin{align}
\frac{H^2}{H_0^2} &= 1 
+ \frac{1}{1-\gamma} \bigg[\Omega_{m,0} \left(a^{-3 \alpha} - 1\right) + \Omega_{r,0} \left(a^{-4 \alpha} - 1\right) 
\bigg], \label{F_1}
\end{align}
where we have used $\Omega_{m,0} +\Omega_{r,0} + \Omega_{e,0} = 1$. 
\subsection*{Model - II}
\noindent
In the second scenario, we examine a situation in which energy is transferred from radiation and baryons to dark matter, in addition to the adiabatic creation of dark matter. Additionally, we postulate the transfer of energy from dark matter, baryonic matter, and radiation to effective entropic dark energy, thereby establishing an interaction between effective entropic dark energy and all constituents. In this context, we assume that radiation and baryons adhere to the standard continuity equations governed by $\rho_b = \rho_{b,0}a^{-3}$ and $\rho_r = \rho_{r,0}a^{-4}$. The continuity equations for $\rho_{dm}$ and $\rho_e$ are given by
\begin{align}    
&\dot\rho_{dm}+3H\left(\rho_{dm}+\frac{p_{eff,dm}}{c^2}\right) = -\dot\rho_e, \label{rho_dm}\\
&\dot\rho_e=-3\gamma H\left((1-\frac{\Gamma_0}{3})\rho_{dm}+\rho_b+\frac{4}{3}\rho_r\right)
\end{align}
where $\dot\rho_e= 3 \gamma H \dot H/(4\pi G)$ is applicable under this scenario, with
$\dot H = -4\pi G\left((1-\frac{\Gamma_0}{3})\rho_{dm}+\rho_b+\frac{4}{3}\rho_r\right).$
We can easily find the exact solutions for $\rho_{dm}$ and $\rho_e$, by solving the above differential equations and the solution for dark matter $\rho_{dm}$ reads as
\begin{align}
\rho_{dm} = &
 \rho_{dm,0} 
        a^{-3\alpha}
    +\frac{\gamma}{\alpha-1}\rho_{b,0}\left(a^{-3}-a^{-3\alpha}\right)\nonumber\\
    &+\frac{4\gamma}{3\alpha-4}\rho_{r,0}\left(a^{-4}-a^{-3\alpha}\right).
\end{align}
and for the effective entropic density $\rho_e$, it gives
\begin{align}
\rho_e = & \rho_{e,0} 
+ \frac{\gamma}{1 - \gamma} \bigg[
    \left( 
        a^{-3\alpha} - 1
    \right) \rho_{\text{dm,0}} \nonumber \\ 
&  + \left( 
        (1 - \gamma)a^{-3} + 
        \gamma \left(\frac{\alpha a^{-3} -a^{-3\alpha}}{\alpha-1}\right)\ - 1
    \right) \rho_{\text{b,0}} \nonumber\\
&  + \left(
        (1 - \gamma)a^{-4} + 
        \gamma \left(\frac{3\alpha a^{-4}- 4 a^{-3\alpha}}{3\alpha-4}\right) - 1
    \right) \rho_{\text{r,0}}
\bigg].
\end{align}
Furthermore, we write the effective equation of state $w_e$ for entropic dark energy in this scenario as well, which reads as
\be
w_{e}= - 1 + \gamma\left[\frac{(1-\frac{\Gamma_0}{3})\Omega_{dm}(a)}{\Omega_{e}(a)}+\frac{H_0^2}{H^2}\frac{\Omega_{b,0}a^{-3}+\frac{4}{3}\Omega_{r,0}a^{-4}}{\Omega_{e}(a)}\right].
\ee
The normalized Friedmann equation for this case can be written as
\begin{align}
\frac{H^2}{H_0^2} &= 1+\frac{1}{1-\gamma} \bigg[ 
\Omega_{dm,0}\left(a^{-3\alpha} - 1\right) \nonumber \\
&\quad + \Omega_{b,0}\left(\frac{-3\gamma a^{-3\alpha} - (1-\gamma)\Gamma_0 a^{-3}}{3(\alpha-1)} - 1\right) \nonumber \\
&\quad + \Omega_{r,0}\bigg(\frac{-4\gamma a^{-3\alpha} }{3\alpha - 4}
 - \frac{(1-\gamma)(1+\Gamma_0)a^{-4}}{3\alpha - 4} - 1\bigg) \bigg],
\end{align}
where we have used $\Omega_{m,}+\Omega_{r,0} + \Omega_{e,0} = 1$ with the definition $\Omega_{m,0}=\Omega_{dm,0} +\Omega_{b,0}$.
\section{Datasets and Methodology}
\noindent
The aforementioned models will be evaluated in comparison with the most recent set of geometrical cosmological data available, to the best of our knowledge. This includes Type Ia Supernovae (SNeIa), 
Cosmic Chronometers (CC); the data from the Gamma Ray Bursts (GRBs); Baryon Acoustic Oscillations (BAO); as well as Cosmic Microwave Background (CMB) distance priors. The following are the details of the datasets:

{\bf SNeIa}. We use distance moduli from 1701 light curves of 1550 spectroscopically confirmed SNeIa from the Pantheon+ sample\footnote{https://github.com/PantheonPlusSH0ES/DataRelease} \cite{Brout:2022vxf}, covering the redshift range $0.001<z<2.26$. The $\chi^2_{SN}$ is defined as $\chi^2_{SN} = \Delta \boldsymbol{\mathcal{\mu}}^{SN} \; \cdot \; \mathbf{C}^{-1}_{SN} \; \cdot \; \Delta  \boldsymbol{\mathcal{\mu}}^{SN}$, where $\Delta\boldsymbol{\mathcal{\mu}} = \mathcal{\mu}_{\rm theo} - \mathcal{\mu}_{\rm obs}$ is the difference between theoretical and observed values of the distance modulus for each SNeIa, and $\mathbf{C}_{SN}$ is the total covariance matrix (statistical and systematic).
The theoretical distance modulus is calculated as
$\mu_{theo}(z_{hel},z_{HD},\boldsymbol{p}) = 25 + 5 \log_{10} [ d_{L}(z_{hel}, z_{HD}, \boldsymbol{p}) ]$,
where $d_L$ is the luminosity distance (in Mpc),
\begin{equation}
d_L(z_{hel}, z_{HD},\boldsymbol{p})=(1+z_{hel})\int_{0}^{z_{HD}}\frac{c\,dz'}{H(z',\boldsymbol{p})}, 
\end{equation}
with $z_{hel}$, the heliocentric redshift; $z_{HD}$, the Hubble diagram redshift \cite{Carr:2021lcj}; and $\boldsymbol{p}$, the vector of cosmological parameters. The observed distance modulus is
$\mu_{obs} = m_{B} - M$, with $m_{B}$ the standardized SNeIa blue apparent magnitude and $M$ the fiducial absolute magnitude calibrated by using primary distance anchors such as Cepheids. 
Generally, $H_0$ and the absolute magnitude $M$ are degenerate with SNeIa alone, but the Pantheon+ sample includes $77$ SNeIa in host galaxies where the distance moduli can be anchored by Cepheids, breaking the degeneracy and allowing separate constraints on $H_0$ and $M$.
Thus, the vector $\Delta\boldsymbol{\mathcal{\mu}}$ is \begin{equation}
\Delta\boldsymbol{\mathcal{\mu}} = \left\{
  ,\begin{array}{ll}
    m_{B,i} - M - \mu_{Ceph,i} & \hbox{$i \in$ Cepheid hosts} \\
    m_{B,i} - M - \mu_{theo,i} & \hbox{otherwise,}
  \end{array}
\right.
\end{equation} 
with $\mu_{Ceph}$ as the Cepheid-calibrated host-galaxy distance from the Pantheon+ team. We designate this dataset by $Pantheon+\&SH0ES$. Additionally, we employ SNIa exclusively, excluding 77 SNeIa in host galaxies where the distance moduli can be anchored via Cepheids; this dataset is designated by $Pantheon+$. In this scenario, $H_0$ and absolute magnitude exhibit degeneracy, necessitating marginalization over these quantities as outlined in \cite{Conley:2011ku}. The $\chi^2$ for this can be expressed as
\begin{equation}
\chi^2_{SN} = a + \log\left( \frac{d}{2\pi} \right) - \frac{b^2}{d}, \tag{59}
\end{equation}
where 
$a \equiv (\Delta \mu_{SN})^{T} \cdot C_{SN}^{-1} \cdot \Delta \mu_{SN}$,
$b \equiv (\Delta\mu_{SN})^{T} \cdot C_{SN}^{-1} \cdot \mathbf{1}$,
$d \equiv \mathbf{1}^{T} \cdot C_{SN}^{-1} \cdot \mathbf{1}$,
and $\mathbf{1}$ is the identity matrix.

{\bf CC.} We use the measurement of Hubble parameter from cosmic chronometers \cite{Jimenez:2001gg,Moresco:2010wh,Moresco:2018xdr,Moresco:2020fbm,Moresco:2022phi}, early-type galaxies which undergo passive evolution and have a characteristic feature in their spectra, 
from which the Hubble parameter $H(z)$ can be measured \cite{Moresco:2012by,Moresco:2012jh,Moresco:2015cya,Moresco:2016nqq,Moresco:2017hwt,Jimenez:2019onw,Jiao:2022aep}. The most updated sample is from \cite{Jiao:2022aep} and spans the redshift range $0<z<1.965$.  
$\chi^2_{H}$ is written as 
$\chi^2_ {H} = \Delta \boldsymbol{\mathcal{H}} \; \cdot \; \mathbf{C}^{-1}_{H} \; \cdot \; \Delta  \boldsymbol{\mathcal{H}}$
where $\Delta \boldsymbol{\mathcal{H}} = H_{theo} - H_{data}$ is the difference between the theoretical and observed Hubble parameter and $\mathbf{C}_{H}$ is the total covariance matrix calculated following \cite{Moresco:2020fbm}.

{\bf GRBs.} We incorporate the measurement of the Hubble parameter from the gamma-ray bursts "Mayflower" sample\footnote{see Table 3 of \cite{Liu:2014vda}} \cite{Liu:2014vda}, which has been calibrated in a cosmological model-independent manner and spans the specified redshift interval $1.44<z<8.1$. The $\chi_{G}^2$ is defined similarly to that for SNeIa; however, we are unable to distinguish between the $H_0$ and the absolute magnitude, necessitating a marginalization over these quantities, in accordance with \cite{Conley:2011ku}.

{\bf CMB.}
The cosmic microwave background is examined through a compressed likelihood methodology, which utilizes the shift parameters, $R(\boldsymbol{p}) \equiv \sqrt{\Omega_m H^2_{0}} r(z_{\ast},\boldsymbol{p})/c$ and 
$l_{a}(\boldsymbol{p}) \equiv \pi r(z_{\ast},\boldsymbol{p})/r_{s}(z_{\ast},\boldsymbol{p})$, originally established in \cite{Wang:2007mza} and subsequently updated in \cite{Zhai:2019nad}, in accordance with the most recent data release presented in \textit{Planck} $2018$ \cite{Planck:2018vyg}. Furthermore, the $\chi^2_{CMB}$ is defined as $\chi^2_{CMB} = \Delta \boldsymbol{\mathcal{F}}^{CMB} \; \cdot \; \mathbf{C}^{-1}_{CMB} \; \cdot \; \Delta  \boldsymbol{\mathcal{F}}^{CMB}$, where the vector $\mathcal{F}^{CMB}$ corresponds to specific quantities in addition to the constraints on baryonic matter content $\Omega_b\,h^2$ and dark matter content $(\Omega_{m,0}-\Omega_{b,0})h^2$. The photon-decoupling redshift $z_{\ast}$ \cite{Hu:1995en} is determined via the most recent fitting formula from \cite{Aizpuru:2021vhd}. The comoving distance at decoupling, denoted as $r(z_{\ast}, \boldsymbol{p})$, is calculated utilizing the definition of comoving distance by setting $r(z_{\ast},\boldsymbol{p}) = D_M(z_{\ast},\boldsymbol{p})=\int_{0}^{z} \frac{c\, dz'}{H(z',\boldsymbol{p})}$. Additionally, $r_{s}(z,\boldsymbol{p}) = \int^{\infty}_{z} \frac{c_{s}(z')}{H(z',\boldsymbol{p})}
\mathrm{d}z'$ represents the comoving sound horizon at the photon-decoupling redshift $z_{\star}$, where the sound speed, $c_{s}(z) = c/\sqrt{3(1+\overline{R}_{b}\, (1+z)^{-1})}$ is provided , and the baryon-to-photon density ratio parameter is specified as $\overline{R}_{b}= 31500 \Omega_{b} \, h^{2} \left( T_{CMB}/ 2.7 \right)^{-4}$ and $T_{CMB} = 2.726$ K.
The sound speed, $c_{s}(z)$, requires a generalization as it is valid solely under the conditions that baryons scale $\propto a^{-3}$ and radiation $\propto a^{-4}$, implying that their respective equations of state parameters $w_i$ are $0$ and $1/3$. However, in the context of Model-I, the continuity equations are altered via the \textit{effective} equations of state parameters, $w_{eff,i}$, which may deviate from typical values. The baryon-to-photon ratio is defined as
$R_b \equiv \frac{\rho_b + p_b}{\rho_\gamma + p_\gamma} = \frac{\rho_b(1+w_b)}{\rho_\gamma(1+w_\gamma)}
$
which, if $w_b=0$ and $w_\gamma=1/3$, becomes the standard $R_b$
with $\overline{R}_{b}$ and its numerical factors coming from  $\Omega_\gamma = \Omega_{r}/(1+0.2271\,N_{eff}) \approx 2.469 \cdot 10^{-5} h^{-2}$ \cite{WMAP:2008lyn}. 
For Model I, $\rho_b$ and $\rho_r$ do not follow the standard continuity equations. As a result, we have modified the baryon-to-photon ratio, 
\begin{equation}
R_b = \frac{(1+w_{eff,b})}{(1+w_{eff,\gamma})}\frac{\Omega_{b,0}}{\Omega_{\gamma,0}}a^{\alpha}= \frac{3\Omega_{b,0}}{4\Omega_{\gamma,0}}(1+z)^{-\alpha} \, .
\end{equation}
Naturally, early-time physics is substantially and robustly constrained by a multitude of observational probes, and it is recognized that modifications within this regime could potentially result in highly unconventional outcomes. However, our approach transcends mere theoretical qualitative assertions; we intend to empirically evaluate these equations through direct data analysis. Consequently, any constraints that emerge will be statistically interpreted concerning their consistency with the equations.

{\bf BAO.} For the analysis of baryon acoustic oscillations, we utilize data sets derived from the latest BAO measurements conducted through the dark-energy spectroscopic instrument (DESI-DR2)\cite{DESI:2025zgx}. These measurements are based on data from Bright Galaxies, Luminous Red Galaxies, Emission Line Galaxies, Quasars and Lyman-$\alpha$
Forest, as shown in Table IV of Ref. \cite{DESI:2025zgx}. They cover an effective redshift range from approximately $z \approx 0.1\text{-} 4.2$. The analyses consider the transverse comoving distance $D_{M}(z,\boldsymbol{p}) / r_{s}(z_{d},\boldsymbol{p})$, the Hubble distance $D_H/r_{s}(z_{d},\boldsymbol{p})$, and the angle-averaged distance $D_V /r_{s}(z_{d},\boldsymbol{p})$, each standardized to the comoving sound horizon at the drag epoch, $r_s$. The correlations between the measurements $D_M/r_s$ and $D_H/r_s$ are taken into account in the calculations. The drag epoch redshift $z_{d}$ \cite{ Hu:1995en} is determined using the latest fitting formulae from \cite{Aizpuru:2021vhd}.
The $\chi_{BAO}^2$ is defined as
$\chi^2_{BAO} = \Delta \boldsymbol{\mathcal{F}}^{BAO} \, \cdot \ \mathbf{C}^{-1}_{BAO} \, \cdot \, \Delta  \boldsymbol{\mathcal{F}}^{BAO}$
with the observables $\mathcal{F}^{BAO}$ corresponds to $D_{M}(z,\boldsymbol{p}) / r_{s}(z_{d},\boldsymbol{p})$, $D_H/r_{s}(z_{d},\boldsymbol{p})$, and $D_V /r_{s}(z_{d},\boldsymbol{p})$ with
$D_{V}(z,\boldsymbol{p})=\left[c z (1+z)^2 D^{2}_{A}(z,\boldsymbol{p})/H(z,\boldsymbol{p})\right]^{1/3}$ and $D_{A}$ is the angular diameter distance defined as $D_{A} = D_{M}/(1+z)$.

The total $\chi^2$ is minimized using the publicly available sampler \cite{emcee}, \texttt{emcee}\footnote{https://github.com/dfm/emcee}, which employs a pure-Python implementation of the affine invariant Markov chain Monte Carlo (MCMC) method. To evaluate the convergence and sampling efficiency of our MCMC analysis, we adhere to the recommendations given in \cite{emcee, zbMATH05709093}. Specifically, we employ the integrated autocorrelation time ($\tau$) to estimate the effective number of independent samples within our chains. This metric considers the correlation among successive samples in a Markov chain and provides a quantitative assessment of the Monte Carlo error in estimated quantities. Furthermore, we utilize trace plots to visually inspect the stability and mixing of walkers, and evaluate the acceptance fraction to determine sampler efficiency, ensuring it falls within the ideal range $0.2\text{--}0.5$. 

We adopt uniform priors on the model parameters within physically motivated bounds: \( 0 < \Omega_{b,0} < \Omega_{m,0} < 1 \), \( 0 < h < 1 \), \( M < 0 \), \( 0 < \gamma < 1 \), and \( -10 < \Gamma_0 < 10 \). To sample the posterior distribution, we use the number of walkers, \( n_{\text{walkers}} = 50 \), chosen to be at least three times the number of model dimensions. The chains are initialized with a burn-in phase of $3000$ steps, followed by $10000$ steps used for the final analysis. To improve sampling efficiency and maintain a suitable acceptance fraction, we use a mixture of proposal moves: for example, a mixture of \texttt{StretchMove}, \texttt{DEMove}, and \texttt{KDEMove}. For more details, see \cite{emcee} and official documentation of \texttt{emcee}.

We use the publicly available \textsc{MCEvidence} code to compute the Bayesian evidence directly from our MCMC chains, enabling quantitative model comparison without additional likelihood evaluations~\cite{Heavens:2017afp}.
To compare two models, $\mathcal{M}_1$ and $\mathcal{M}_2$, we compute the \textit{Bayes factor}:
$B_{12} = \frac{\mathcal{Z}_2}{\mathcal{Z}_1}$, where the \textit{Bayesian evidence}, denoted as $\mathcal{Z}$.
For practical interpretation, we use the natural logarithm of the Bayes factor:
\begin{equation}
\ln B_{1,2} = \ln \mathcal{Z}_2 - \ln \mathcal{Z}_1.
\end{equation}
We interpret $\ln B_{12}$ using the Jeffreys scale \cite{Jeffreys:1961}. Negative values of $\ln B_{1,2}$ indicate evidence in favor of model $\mathcal{M}_1$, while positive values indicate evidence in favor of model $\mathcal{M}_2$.
This provides a statistically robust and widely accepted framework for testing extensions to the standard $\Lambda$CDM model. For the comparison, we consider $\mathcal{M}_1 = \Lambda$CDM as the reference model.

Moreover, we also employ the Akaike Information Criterion (AIC) and the Deviance Information Criterion (DIC), both of which quantify the trade-off between goodness of fit and model complexity. The AIC~\cite{Akaike1974} is defined as
\begin{equation}
\mathrm{AIC} = \chi^2_{\min} + \frac{2nN}{N - n - 1},
\end{equation}
where $\chi^2_{\min}$ is the minimum chi-square value, $n$ is the number of free parameters, and $N$ is the number of data points. The DIC~\cite{Spiegelhalter2002} is a Bayesian generalization based on the deviance $D(\theta) = -2\ln \mathcal{L}(\theta)$,
\begin{equation}
\mathrm{DIC} = \overline{D} + p_D = 2\overline{D} - D(\hat{\theta}),
\end{equation}
where $\overline{D}$ is the posterior mean of the deviance, $D(\hat{\theta})$ is its value at the posterior mean of the parameters, and $p_D = \overline{D} - D(\hat{\theta})$ represents the effective number of parameters. In both cases, smaller values of the criterion indicate a more favorable model. To compare with the reference $\Lambda$CDM scenario, we define the differences $\Delta\mathrm{AIC} = \mathrm{AIC}_{\Lambda\mathrm{CDM}} - \mathrm{AIC}_{\text{model}}$ and $\Delta\mathrm{DIC} = \mathrm{DIC}_{\Lambda\mathrm{CDM}} - \mathrm{DIC}_{\text{model}}$. Positive values of these differences indicate that the alternative model is statistically preferred over $\Lambda$CDM after penalizing for additional parameters, while negative values imply that $\Lambda$CDM remains favored. According to standard criteria~\cite{KassRaftery1995,BurnhamAnderson2002}, values of $|\Delta|<2$ suggest statistical consistency between models, $2<|\Delta|<6$ indicate positive evidence, $6<|\Delta|<10$ denote strong evidence, and $|\Delta|>10$ correspond to very strong evidence in favor of the model with smaller AIC or DIC.

Furthermore,
we employ the publicly accessible Python package \cite{Lewis:2019xzd} \texttt{GetDist}\footnote{https://github.com/cmbant/getdist} to generate plots and perform statistical analysis.
\begin{table*}[t]
    \centering
    \footnotesize
    \renewcommand{\arraystretch}{1.2}
    \setlength{\tabcolsep}{12pt}
    \begin{tabular}{l c c c}
        \hline
        \multicolumn{4}{c}{Pantheon$+$\&SH0ES+CMB+BAO+GRB+CC}\\
        \hline\hline
        Parameter & Model-I & Model-II & $\Lambda$CDM \\
        \hline
        $\Omega_m$ & $0.2856\pm 0.0044$ & $0.2830\pm 0.0047$ & $0.2934\pm 0.0037$ \\
        $\Omega_b$ & $0.04430\pm 0.00084$ & $0.04509\pm 0.00083$ & $0.04721\pm 0.00042$ \\
        $h$ & $0.7175\pm 0.0079$ & $0.7106\pm 0.0081$ & $0.6887\pm 0.0030$ \\
        $\gamma$ & $0.0128^{+0.0050}_{-0.012}$ & $0.00241^{+0.00084}_{-0.0024}$ & -- \\
        $\Gamma_0$ & $-0.043^{+0.037}_{-0.015}$ & $-0.0211^{+0.011}_{-0.0049}$ & -- \\
        $M$ & $-19.312\pm 0.023$ & $-19.334\pm 0.023$ & $-19.3982\pm 0.0090$ \\
        \hline\hline
        $\ln{\mathcal{Z}_i}$ & $-833.68 \pm 0.14$ & $-837.23 \pm 0.067$ & $-831.24 \pm 0.007$ \\
        $\ln{B}_{\Lambda\text{CDM},i}$ & $-2.43$ & $-5.99$ & $0$ \\
        $\chi^2_{\text{min}}$ & $1601.34$ & $1607.12$ & $1616.56$ \\
        $\Delta$AIC & $11.19$ & $5.32$ & $0$ \\
        $\Delta$DIC & $13.73$ & $5.86$ & $0$ \\
        \hline
    \end{tabular}
    \caption{Constraints with $\gamma > 0$ and $\Gamma_0$ free using Pantheon$+$\&SH0ES+CMB+BAO+GRB+CC data.}
    \label{Table_1}
\end{table*}

\begin{table*}[t]
    \centering
    \footnotesize
    \renewcommand{\arraystretch}{1.2}
    \setlength{\tabcolsep}{12pt}
    \begin{tabular}{l c c c}
        \hline
        \multicolumn{4}{c}{$\Gamma(t)=0$ and Pantheon$+$\&SH0ES+CMB+BAO+GRB+CC} \\
        \hline\hline
        Parameter & Model-I & Model-II & $\Lambda$CDM \\
        \hline
        $\Omega_m$ & $0.2936\pm 0.0036$ & $0.2940\pm 0.0037$ & $0.2934\pm 0.0037$ \\
        $\Omega_b$ & $0.04739\pm 0.00045$ & $0.04740\pm 0.00046$ & $0.04721\pm 0.00042$ \\
        $h$ & $0.6871\pm 0.0033$ & $0.6869\pm 0.0035$ & $0.6887\pm 0.0030$ \\
        $\gamma$ & $0.000093^{+0.000031}_{-0.000095}$ & $0.000245^{+0.000084}_{-0.00025}$ & -- \\
        $M$ & $-19.403\pm 0.010$ & $-19.403\pm 0.010$ & $-19.3982\pm 0.0090$ \\
        \hline\hline
        $\ln{\mathcal{Z}_i}$ & $-840.78 \pm 0.095$ & $-839.71 \pm 0.093$ & $-831.37 \pm 0.042$ \\
        $\ln{B}_{\Lambda\text{CDM},i}$ & $-9.41$ & $-8.34$ & $0$ \\
        $\chi^2_{\text{min}}$ & $1616.79$ & $1616.50$ & $1616.56$ \\
        $\Delta$AIC & $-2.24$ & $-1.95$ & $0$ \\
        $\Delta$DIC & $-1.98$ & $-1.77$ & $0$ \\
        \hline
    \end{tabular}
    \caption{Constraints with $\Gamma(t)=0$ using Pantheon$+$\&SH0ES+CMB+BAO+GRB+CC data.}
    \label{Table_3}
\end{table*}

\begin{table*}[t]
    \centering
    \footnotesize
    \renewcommand{\arraystretch}{1.2}
    \setlength{\tabcolsep}{12pt}
    \begin{tabular}{l c c c}
        \hline
        \multicolumn{4}{c}{$\gamma=0$, $\Gamma_0<0$ and Pantheon$+$\&SH0ES+CMB+BAO+GRB+CC} \\
        \hline\hline
        Parameter & Model-I & Model-II & $\Lambda$CDM \\
        \hline
        $\Omega_m$ & $0.2882\pm 0.0038$ & $0.2843\pm 0.0045$ & $0.2934\pm 0.0037$ \\
        $\Omega_b$ & $0.04422\pm 0.00083$ & $0.04498\pm 0.00081$ & $0.04721\pm 0.00042$ \\
        $h$ & $0.7163\pm 0.0078$ & $0.7110\pm 0.0079$ & $0.6887\pm 0.0030$ \\
        $\Gamma_0$ & $-0.0046\pm 0.0012$ & $-0.0114\pm 0.0035$ & -- \\
        $M$ & $-19.316\pm 0.023$ & $-19.333\pm 0.023$ & $-19.3982\pm 0.0090$ \\
        \hline\hline
        $\ln{\mathcal{Z}_i}$ & $-829.72 \pm 0.069$ & $-831.34 \pm 0.064$ & $-831.37 \pm 0.041$ \\
        $\ln{B}_{\Lambda\text{CDM},i}$ & $1.65$ & $-0.031$ & $0$ \\
        $\chi^2_{\text{min}}$ & $1601.24$ & $1606.98$ & $1616.56$ \\
        $\Delta$AIC & $13.31$ & $7.57$ & $0$ \\
        $\Delta$DIC & $13.34$ & $7.72$ & $0$ \\
        \hline
    \end{tabular}
    \caption{Constraints with $\gamma=0$ and $\Gamma_0<0$ using Pantheon$+$\&SH0ES+CMB+BAO+GRB+CC data.}
    \label{Table_5}
\end{table*}

\begin{table*}[t]
    \centering
    \footnotesize
    \renewcommand{\arraystretch}{1.2}
    \setlength{\tabcolsep}{12pt}
    \begin{tabular}{l c c c}
        \hline
        \multicolumn{4}{c}{$\gamma=0$, $\Gamma_0>0$ and Pantheon$+$\&SH0ES+CMB+BAO+GRB+CC} \\
        \hline\hline
        Parameter & Model-I & Model-II & $\Lambda$CDM \\
        \hline
        $\Omega_m$ & $0.2937\pm 0.0037$ & $0.2942\pm 0.0037$ & $0.2934\pm 0.0037$ \\
        $\Omega_b$ & $0.04739\pm 0.00045$ & $0.04739\pm 0.00046$ & $0.04721\pm 0.00042$ \\
        $h$ & $0.6870^{+0.0035}_{-0.0031}$ & $0.6868^{+0.0037}_{-0.0032}$ & $0.6887\pm 0.0030$ \\
        $\Gamma_0$ & $0.000274^{+0.000091}_{-0.00028}$ & $0.00110^{+0.00037}_{-0.0011}$ & -- \\
        $M$ & $-19.403^{+0.010}_{-0.0093}$ & $-19.404^{+0.011}_{-0.0097}$ & $-19.3982\pm 0.0090$ \\
        \hline\hline
        $\ln{\mathcal{Z}_i}$ & $-839.70 \pm 0.094$ & $-838.22 \pm 0.093$ & $-831.37 \pm 0.041$ \\
        $\ln{B}_{\Lambda\text{CDM},i}$ & $-8.33$ & $-6.85$ & $0$ \\
        $\chi^2_{\text{min}}$ & $1616.71$ & $1616.58$ & $1616.56$ \\
        $\Delta$AIC & $-2.16$ & $-2.03$ & $0$ \\
        $\Delta$DIC & $-1.97$ & $-1.82$ & $0$ \\
        \hline
    \end{tabular}
    \caption{Constraints with $\gamma=0$ and $\Gamma_0>0$ using Pantheon$+$\&SH0ES+CMB+BAO+GRB+CC data.}
    \label{Table_6}
\end{table*}

\begin{table*}[t]
    \centering
    \footnotesize
    \renewcommand{\arraystretch}{1.2}
    \setlength{\tabcolsep}{12pt}
    \begin{tabular}{l c c c}
        \hline
        \multicolumn{4}{c}{Pantheon$+$+CMB+BAO+GRB+CC} \\
        \hline\hline
        Parameter & Model-I & Model-II & $\Lambda$CDM \\
        \hline
        $\Omega_m$ & $0.2972\pm 0.0055$ & $0.3067\pm 0.0082$ & $0.2987\pm 0.0039$ \\
        $\Omega_b$ & $0.0479\pm 0.0015$ & $0.0496\pm 0.0014$ & $0.04753\pm 0.00043$ \\
        $h$ & $0.682\pm 0.013$ & $0.665\pm 0.012$ & $0.6838\pm 0.0031$ \\
        $\gamma$ & $0.0095^{+0.0034}_{-0.0094}$ & $0.0033^{+0.0012}_{-0.0033}$ & -- \\
        $\Gamma_0$ & $-0.028^{+0.029}_{-0.010}$ & $-0.0038^{+0.015}_{-0.0080}$ & -- \\
        \hline\hline
        $\ln{\mathcal{Z}_i}$ & $-794.47 \pm 0.13$ & $-793.46 \pm 0.10$ & $-784.50 \pm 0.028$ \\
        $\ln{B}_{\Lambda\text{CDM},i}$ & $-9.97$ & $-8.96$ & $0$ \\
        $\chi^2_{\text{min}}$ & $1532.55$ & $1530.65$ & $1532.52$ \\
        $\Delta$AIC & $-4.06$ & $-2.15$ & $0$ \\
        $\Delta$DIC & $-0.58$ & $-1.89$ & $0$ \\
        \hline
    \end{tabular}
    \caption{Constraints with $\gamma > 0$ and $\Gamma_0$ free using Pantheon$+$+CMB+BAO+GRB+CC data (without SH0ES).}
    \label{Table_2}
\end{table*}

\begin{table*}[t]
    \centering
    \footnotesize
    \renewcommand{\arraystretch}{1.2}
    \setlength{\tabcolsep}{12pt}
    \begin{tabular}{l c c c}
        \hline
        \multicolumn{4}{c}{$\Gamma(t)=0$ and Pantheon$+$+CMB+BAO+GRB+CC} \\
        \hline\hline
        Parameter & Model-I & Model-II & $\Lambda$CDM \\
        \hline
        $\Omega_m$ & $0.3016\pm 0.0044$ & $0.3092^{+0.0038}_{-0.010}$ & $0.2987\pm 0.0039$ \\
        $\Omega_b$ & $0.04877^{+0.00068}_{-0.0012}$ & $0.04986^{+0.00072}_{-0.0020}$ & $0.04753\pm 0.00043$ \\
        $h$ & $0.673^{+0.010}_{-0.0056}$ & $0.664^{+0.016}_{-0.0064}$ & $0.6838\pm 0.0031$ \\
        $\gamma$ & $0.00058^{+0.00023}_{-0.00053}$ & $0.00279^{+0.00076}_{-0.0023}$ & -- \\
        \hline\hline
        $\ln{\mathcal{Z}_i}$ & $-791.48 \pm 0.039$ & $-789.57 \pm 0.10$ & $-784.50 \pm 0.028$ \\
        $\ln{B}_{\Lambda\text{CDM},i}$ & $-6.97$ & $-5.07$ & $0$ \\
        $\chi^2_{\text{min}}$ & $1532.50$ & $1531.36$ & $1532.52$ \\
        $\Delta$AIC & $-1.99$ & $-0.85$ & $0$ \\
        $\Delta$DIC & $-1.23$ & $-3.38$ & $0$ \\
        \hline
    \end{tabular}
    \caption{Constraints with $\Gamma(t)=0$ using Pantheon$+$+CMB+BAO+GRB+CC data (without SH0ES).}
    \label{Table_4}
\end{table*}

\begin{table*}[t]
    \centering
    \footnotesize
    \renewcommand{\arraystretch}{1.2}
    \setlength{\tabcolsep}{11pt}
    \begin{tabular}{l c c c}
        \hline
        \multicolumn{4}{c}{$\gamma=0$, $\Gamma_0<0$ and Pantheon$+$+CMB+BAO+GRB+CC} \\
        \hline\hline
        Parameter & Model-I & Model-II & $\Lambda$CDM \\
        \hline
        $\Omega_m$ & $0.2963\pm 0.0042$ & $0.2960\pm 0.0045$ & $0.2987\pm 0.0039$ \\
        $\Omega_b$ & $0.04657^{+0.00099}_{-0.00059}$ & $0.04697^{+0.00069}_{-0.00049}$ & $0.04753\pm 0.00043$ \\
        $h$ & $0.6927^{+0.0048}_{-0.0091}$ & $0.6894^{+0.0038}_{-0.0065}$ & $0.6838\pm 0.0031$ \\
        $\Gamma_0$ & $-0.00134^{+0.0013}_{-0.00052}$ & $-0.00261^{+0.0026}_{-0.00095}$ & -- \\
        \hline\hline
        $\ln{\mathcal{Z}_i}$ & $-790.89 \pm 0.065$ & $-790.56 \pm 0.099$ & $-784.57 \pm 0.045$ \\
        $\ln{B}_{\Lambda\text{CDM},i}$ & $-6.31$ & $-5.99$ & $0$ \\
        $\chi^2_{\text{min}}$ & $1532.53$ & $1532.53$ & $1532.52$ \\
        $\Delta$AIC & $-2.02$ & $-2.02$ & $0$ \\
        $\Delta$DIC & $-1.40$ & $-1.73$ & $0$ \\
        \hline
    \end{tabular}
    \caption{Constraints with $\gamma=0$ and $\Gamma_0<0$ using Pantheon$+$+CMB+BAO+GRB+CC data (without SH0ES).}
    \label{Table_7}
\end{table*}

\begin{table*}[t]
    \centering
    \footnotesize
    \renewcommand{\arraystretch}{1.2}
    \setlength{\tabcolsep}{11pt}
    \begin{tabular}{l c c c}
        \hline
        \multicolumn{4}{c}{$\gamma=0$, $\Gamma_0>0$ and Pantheon$+$+CMB+BAO+GRB+CC} \\
        \hline\hline
        Parameter & Model-I & Model-II & $\Lambda$CDM \\
        \hline
        $\Omega_m$ & $0.3018^{+0.0042}_{-0.0047}$ & $0.3120^{+0.0043}_{-0.011}$ & $0.2987\pm 0.0039$ \\
        $\Omega_b$ & $0.04880^{+0.00071}_{-0.0013}$ & $0.04980^{+0.00075}_{-0.0016}$ & $0.04753\pm 0.00043$ \\
        $h$ & $0.673^{+0.011}_{-0.0058}$ & $0.663^{+0.014}_{-0.0069}$ & $0.6838\pm 0.0031$ \\
        $\Gamma_0$ & $0.00177^{+0.00070}_{-0.0016}$ & $0.0124^{+0.0033}_{-0.0090}$ & -- \\
        \hline\hline
        $\ln{\mathcal{Z}_i}$ & $-790.45 \pm 0.064$ & $-787.85 \pm 0.14$ & $-784.57 \pm 0.045$ \\
        $\ln{B}_{\Lambda\text{CDM},i}$ & $-5.88$ & $-3.28$ & $0$ \\
        $\chi^2_{\text{min}}$ & $1532.49$ & $1530.53$ & $1532.52$ \\
        $\Delta$AIC & $-1.98$ & $-0.02$ & $0$ \\
        $\Delta$DIC & $-1.21$ & $1.43$ & $0$ \\
        \hline
    \end{tabular}
    \caption{Constraints with $\gamma=0$ and $\Gamma_0>0$ using Pantheon$+$+CMB+BAO+GRB+CC data (without SH0ES).}
    \label{Table_8}
\end{table*}

\section{Results and Discussion}
\begin{figure*}[t]
    \centering
    \begin{minipage}[t]{0.48\textwidth}
        \centering
        \includegraphics[width=\textwidth]{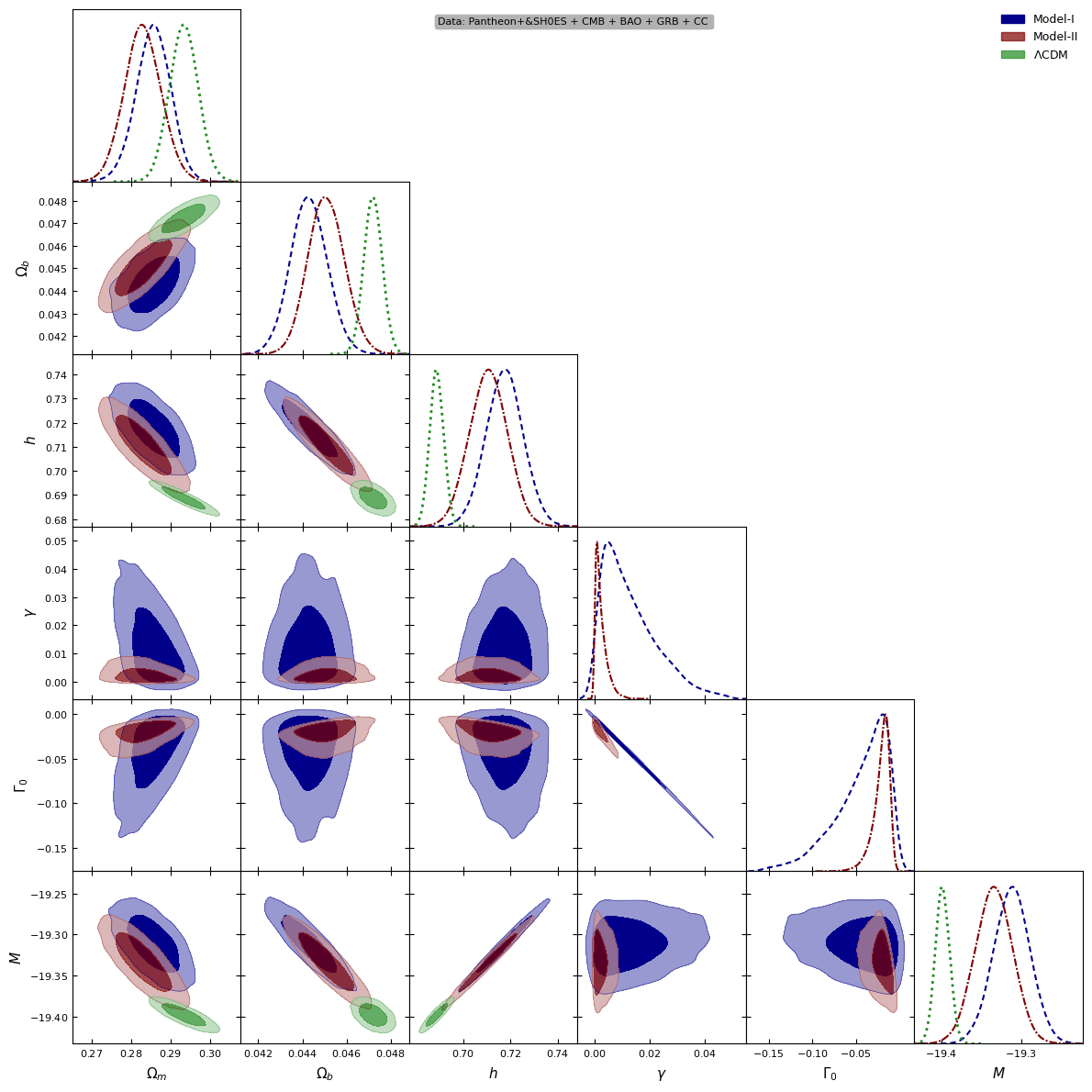}
        \caption{Contour plots representing $68\%$ and $95\%$ confidence regions for Models I, II, and $\Lambda$CDM with Pantheon$+$$\&$SH0ES+CMB+BAO+GRB+CC data. Priors: $\gamma > 0$ and $\Gamma_0$ free.}
        \label{contourPlot_1}
    \end{minipage}
    \hfill
    \begin{minipage}[t]{0.48\textwidth}
        \centering
        \includegraphics[width=\textwidth]{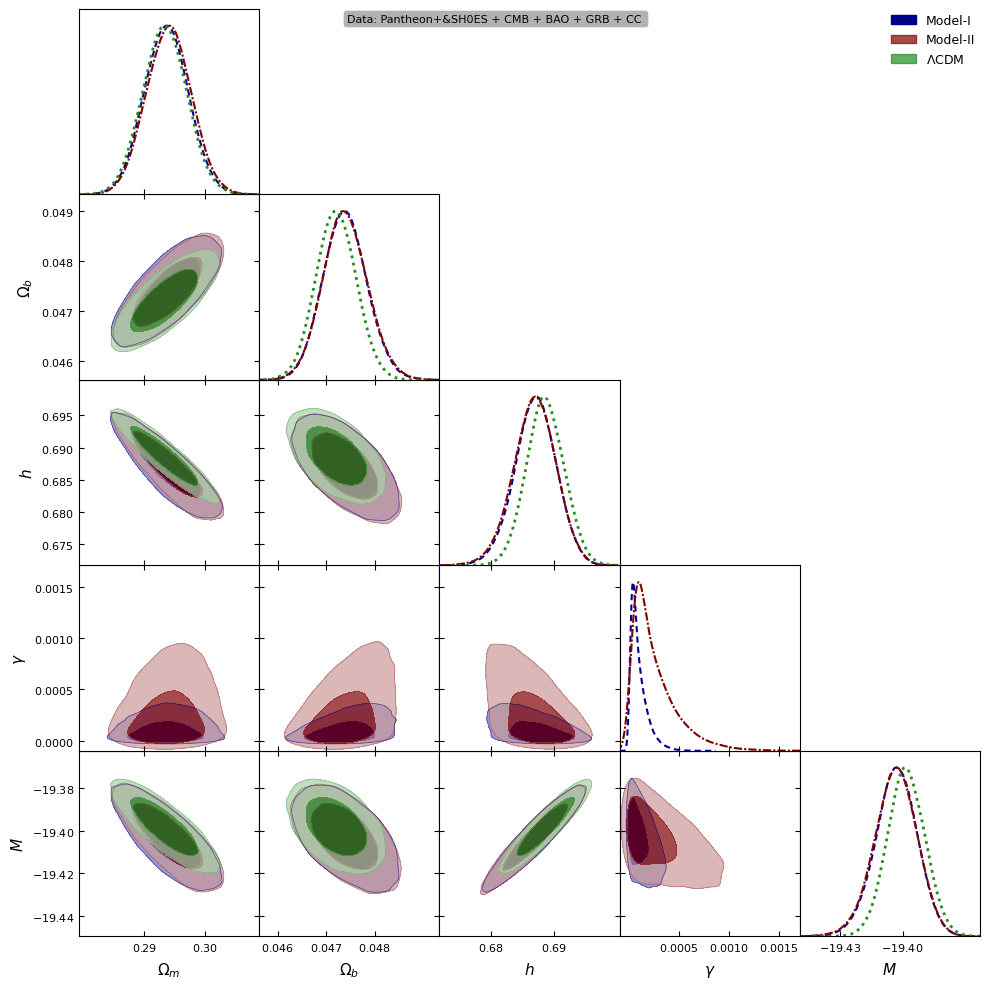}
        \caption{Same as Fig.~\ref{contourPlot_1} but with matter creation/annihilation function $\Gamma(t)=0$.}
        \label{contourPlot_3}
    \end{minipage}
    
    \vspace{0.5cm}
    
    \begin{minipage}[t]{0.48\textwidth}
        \centering
        \includegraphics[width=\textwidth]{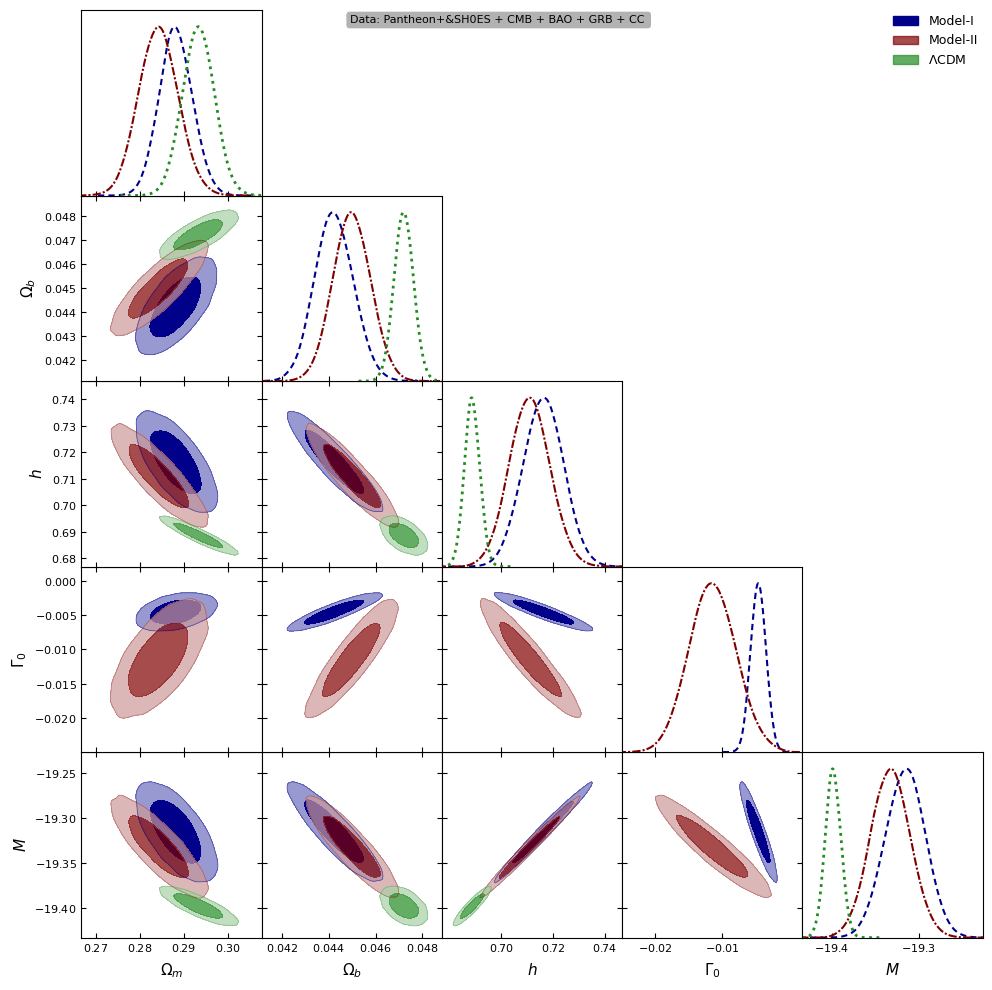}
        \caption{Same as Fig.~\ref{contourPlot_1} but with $\gamma=0$ and prior $\Gamma_0<0$.}
        \label{contourPlot_5}
    \end{minipage}
    \hfill
    \begin{minipage}[t]{0.48\textwidth}
        \centering
        \includegraphics[width=\textwidth]{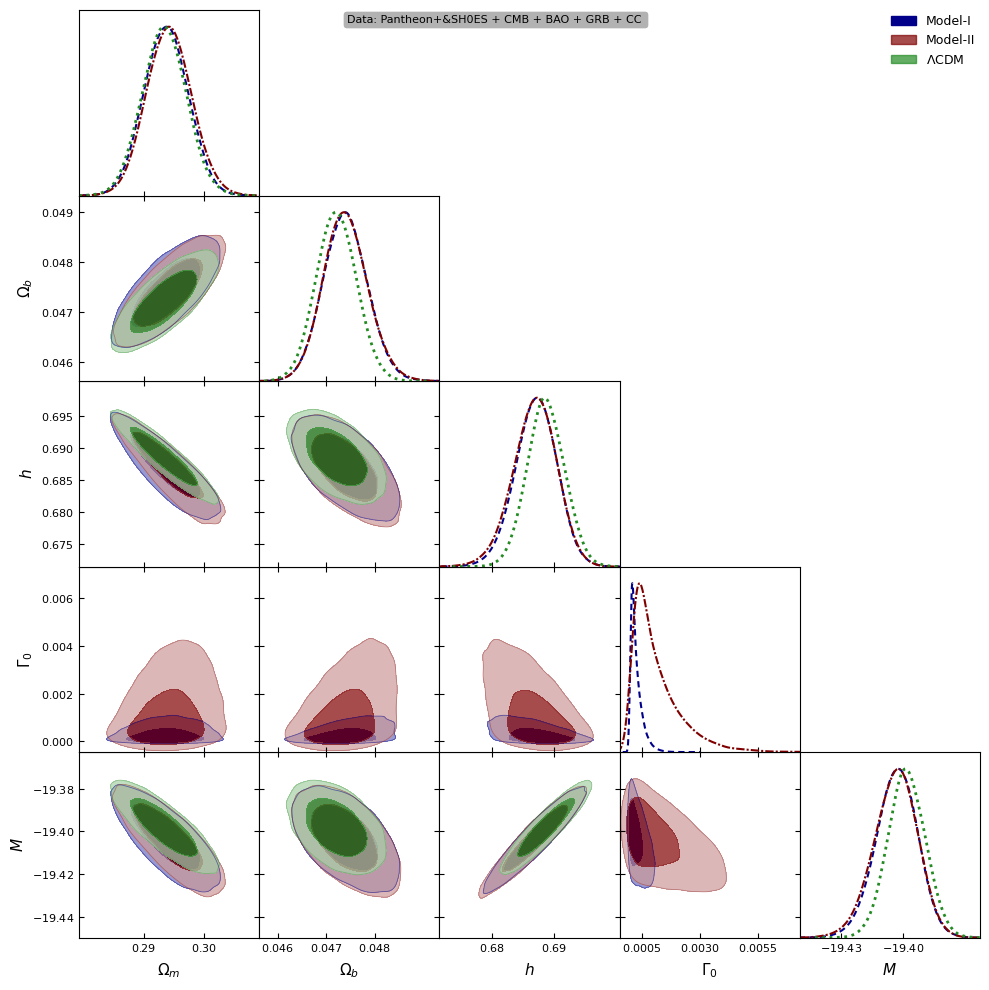}
        \caption{Same as Fig.~\ref{contourPlot_1} but with $\gamma=0$ and prior $\Gamma_0>0$.}
        \label{contourPlot_6}
    \end{minipage}
\end{figure*}

\begin{figure*}[t]
    \centering
    \begin{minipage}[t]{0.48\textwidth}
        \centering
        \includegraphics[width=\textwidth]{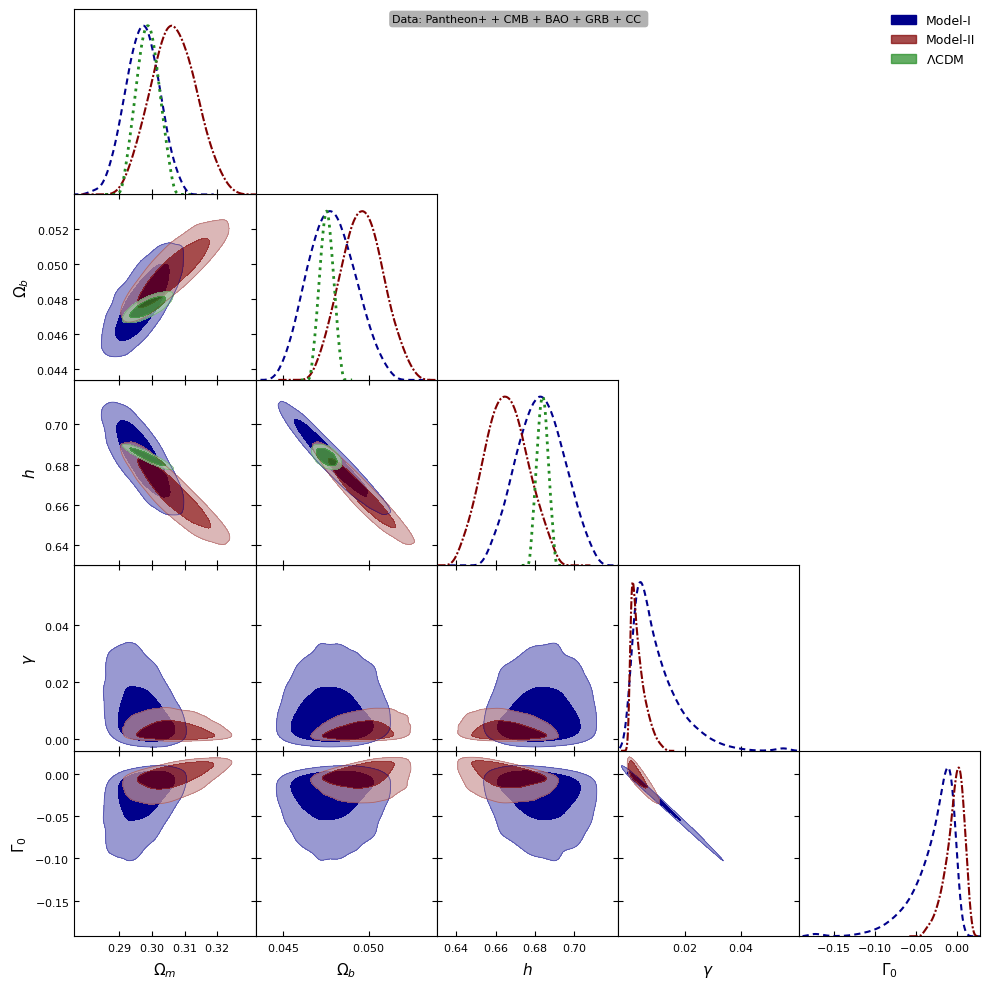}
        \caption{Contour plots representing $68\%$ and $95\%$ confidence regions for Models I, II, and $\Lambda$CDM with Pantheon$+$+CMB+BAO+GRB+CC data (without SH0ES). Priors: $\gamma > 0$ and $\Gamma_0$ free.}
        \label{contourPlot_2}
    \end{minipage}
    \hfill
    \begin{minipage}[t]{0.48\textwidth}
        \centering
        \includegraphics[width=\textwidth]{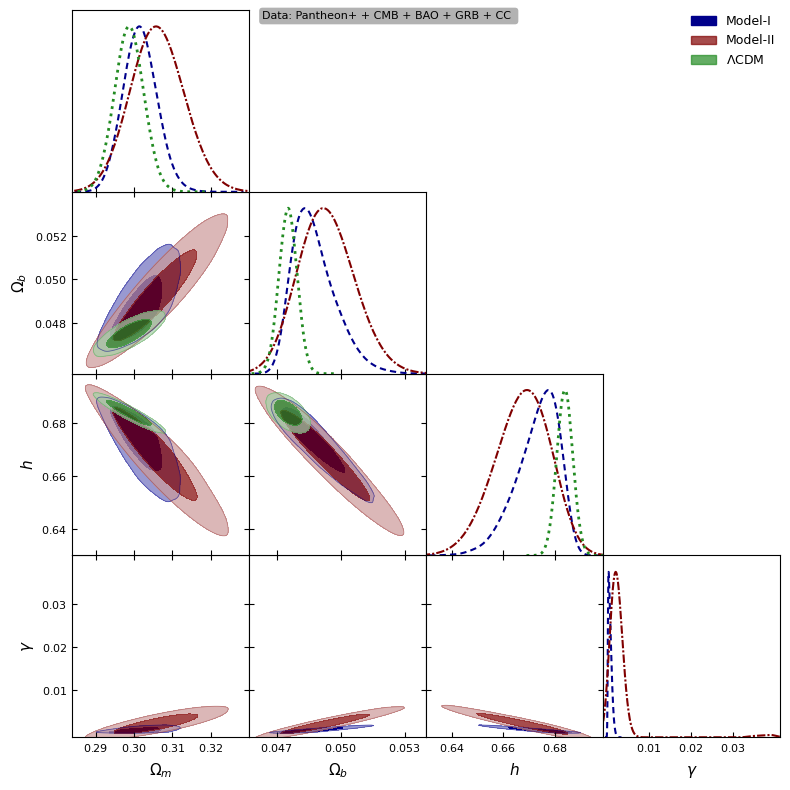}
        \caption{Same as Fig.~\ref{contourPlot_2} but with matter creation/annihilation function $\Gamma(t)=0$.}
        \label{contourPlot_4}
    \end{minipage}
    
    \vspace{0.5cm}
    
    \begin{minipage}[t]{0.48\textwidth}
        \centering
        \includegraphics[width=\textwidth]{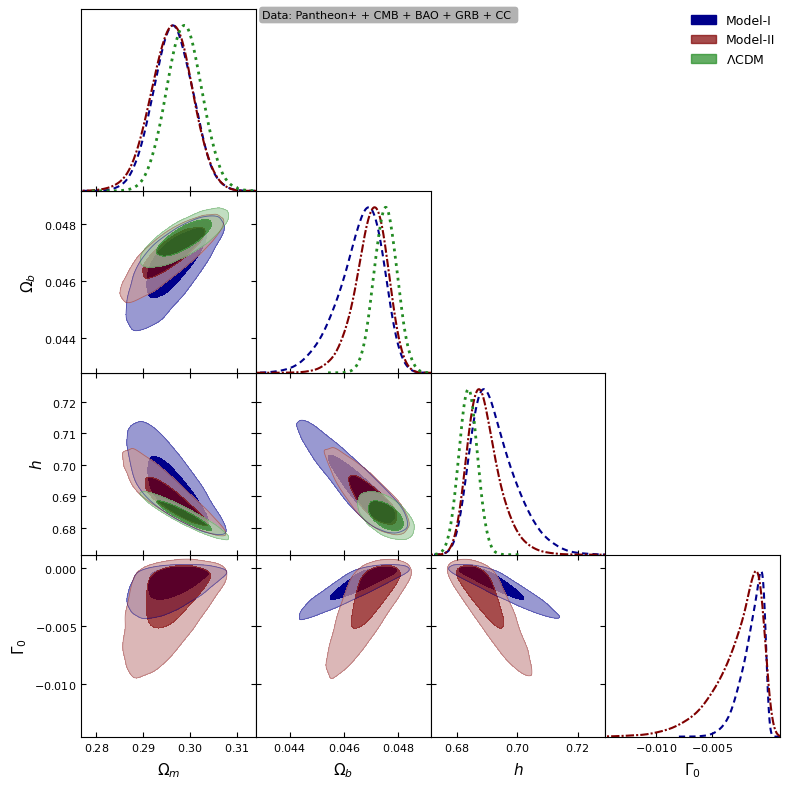}
        \caption{Same as Fig.~\ref{contourPlot_2} but with $\gamma=0$ and prior $\Gamma_0<0$.}
        \label{contourPlot_7}
    \end{minipage}
    \hfill
    \begin{minipage}[t]{0.48\textwidth}
        \centering
        \includegraphics[width=\textwidth]{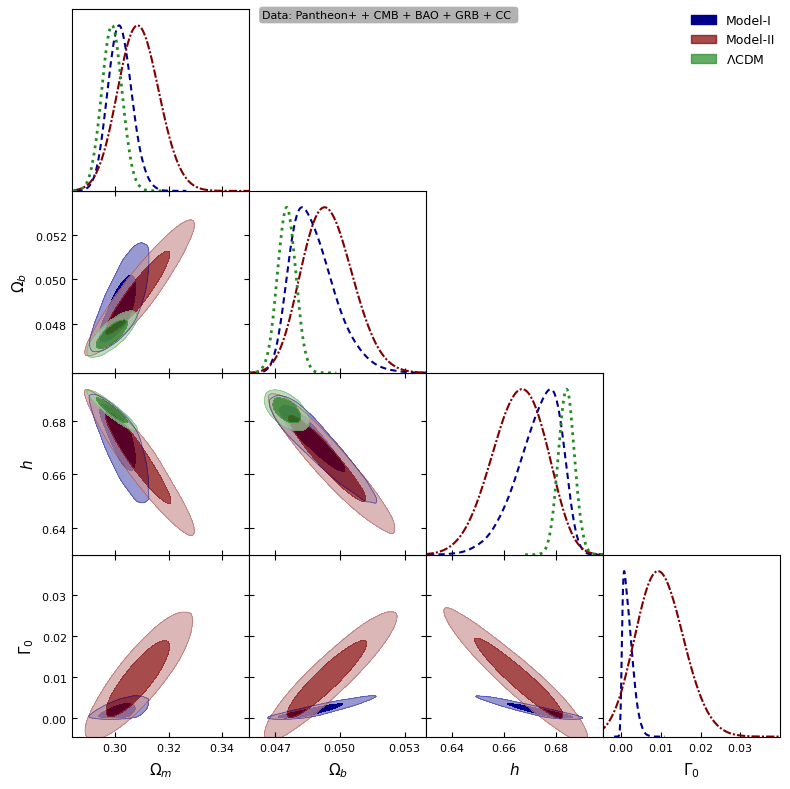}
        \caption{Same as Fig.~\ref{contourPlot_2} but with $\gamma=0$ and prior $\Gamma_0>0$.}
        \label{contourPlot_8}
    \end{minipage}
\end{figure*}

\noindent
The primary findings are systematically presented in Tables \ref{Table_1}, \ref{Table_2}, \ref{Table_3}, \ref{Table_4}, \ref{Table_5}, \ref{Table_6}, \ref{Table_7} and \ref{Table_8}. These tables delineate the marginalized mean values and the $68\%$ confidence level limits for each parameter across three distinct models under varying physical scenarios. This analysis encompasses dataset combinations involving SNIa(Pantheon$+$, both inclusive and exclusive of SH$0$ES data), CMB distance priors, BAO (DESI DR2), CC, and GRB. Correspondingly, Figures \ref{contourPlot_1}, \ref{contourPlot_2}, \ref{contourPlot_3}, \ref{contourPlot_4}, \ref{contourPlot_5}, \ref{contourPlot_6}, \ref{contourPlot_7} and \ref{contourPlot_8}  illustrate the combined contour plots, delineating each model's parameters alongside the parameters of the standard $\Lambda$CDM model.

For the data set Pantheon$+$$\&$SH$0$ES + CMB + BAO + GRB + CC, a comparative analysis of Models I, II, and $\Lambda$CDM reveals distinct parameter trends that elucidate their capacity to alleviate the Hubble tension. Models I and II, which incorporate the additional parameters $\Gamma_0$ and $\gamma$ representing matter creation/annihilation and energy flow, yield higher $H_0$ values compared to $\Lambda$CDM, thereby reducing the tension with the latest SH0ES measurements reported in 2024 \cite{Breuval:2024lsv} to a range of $1.2$--$1.8\sigma$. Notably, when considering SH0ES measurements from 2021 and 2022 \cite{Riess:2020fzl,Riess:2021jrx}, these models further diminish the tension to approximately $1\sigma$. This mitigation is accomplished through lower matter density ($\Omega_m \sim 0.28$ vs. $0.29$ in $\Lambda$CDM) and reduced baryon density ($\Omega_b \sim 0.044$--$0.045$ vs. $0.047$ in $\Lambda$CDM).

Model I demonstrates a marginal advantage over Model II in alleviating the Hubble tension. Its value, $h = 0.7175 \pm 0.0079$, aligns more closely with the latest SH0ES measurement, $h = 0.7317 \pm 0.0086$, reducing the discrepancy to $\sim1.2\sigma$, while Model II ($h = 0.7106 \pm 0.0081$) maintains approximately $1.8\sigma$ tension. The more pronounced negative value of $\Gamma_0$ in Model I ($\Gamma_0 = -0.043^{+0.037}_{-0.015}$) relative to Model II ($\Gamma_0 = -0.0211^{+0.011}_{-0.0049}$) implies more substantial modifications to late-time cosmic expansion. This is accompanied by a larger value of $\gamma$ in Model I ($\gamma = 0.0128^{+0.0050}_{-0.012}$) compared to Model II ($\gamma = 0.00241^{+0.00084}_{-0.0024}$), which quantifies energy transfer toward the effective entropic dark energy component.

To elucidate the individual contributions of $\gamma$ and $\Gamma_0$, we examine constrained scenarios. When setting $\gamma = 0$ with prior $\Gamma_0 < 0$ (Table \ref{Table_5}), both models successfully alleviate the Hubble tension, with Model I achieving $h = 0.7163 \pm 0.0078$ and Model II yielding $h = 0.7110 \pm 0.0079$. Notably, the magnitude of $\Gamma_0$ is significantly reduced when $\gamma = 0$: in Model I, $\Gamma_0$ shifts from $-0.043^{+0.037}_{-0.015}$ (with $\gamma \neq 0$) to $-0.0046 \pm 0.0012$ (with $\gamma = 0$), representing an approximately 9-fold reduction in magnitude. Similarly, in Model II, $\Gamma_0$ changes from $-0.0211^{+0.011}_{-0.0049}$ to $-0.0114 \pm 0.0035$, roughly a 2-fold reduction. This substantial decrease indicates a strong degeneracy between $\gamma$ and $\Gamma_0$: when energy transfer ($\gamma$) is absent, less matter annihilation ($\Gamma_0$) is required to achieve the same cosmological effects. Despite these reduced magnitudes, the negative values of $\Gamma_0 = -0.0046 \pm 0.0012$ (Model I) and $\Gamma_0 = -0.0114 \pm 0.0035$ (Model II) demonstrate that matter annihilation alone remains crucial for elevating $H_0$ values. 

Conversely, when $\gamma = 0$ with prior $\Gamma_0 > 0$ (Table \ref{Table_6}), both models fail to alleviate the tension, yielding $h$ values ($h = 0.6870^{+0.0035}_{-0.0031}$ for Model I and $h = 0.6868^{+0.0037}_{-0.0032}$ for Model II) consistent with $\Lambda$CDM. In this scenario, $\Gamma_0$ takes very small positive values: $\Gamma_0 = 0.000274^{+0.000091}_{-0.00028}$ for Model I and $\Gamma_0 = 0.00110^{+0.00037}_{-0.0011}$ for Model II, which are orders of magnitude smaller than their negative counterparts. This indicates that matter creation ($\Gamma_0 > 0$) is ineffective in addressing the Hubble tension, underscoring that the sign and magnitude of $\Gamma_0$ are critical, with matter annihilation ($\Gamma_0 < 0$) being essential for tension mitigation.

To further isolate the role of matter creation/annihilation, we consider the case where $\Gamma(t) = 0$ entirely, allowing only the energy transfer parameter $\gamma$ to set free (Table \ref{Table_3}). In this scenario, both models fail to alleviate the Hubble tension, yielding $h = 0.6871 \pm 0.0033$ for Model I and $h = 0.6869 \pm 0.0035$ for Model II, values that are fully consistent with $\Lambda$CDM ($h = 0.6887 \pm 0.0030$). The constrained values of $\gamma$ are extremely small: $\gamma = 0.000093^{+0.000031}_{-0.000095}$ for Model I and $\gamma = 0.000245^{+0.000084}_{-0.00025}$ for Model II, representing reductions of approximately two orders of magnitude compared to the unconstrained case (where $\gamma \sim 0.01$ for Model I and $\gamma \sim 0.002$ for Model II). Similarly, the cosmological parameters $\Omega_m$ and $\Omega_b$ converge to values nearly identical to $\Lambda$CDM. This result unambiguously demonstrates that energy transfer alone, without accompanying matter creation or annihilation, is insufficient to address the Hubble tension. The comparison between Tables \ref{Table_1}, \ref{Table_5}, and \ref{Table_7} establishes a clear hierarchy: matter annihilation ($\Gamma_0 < 0$) is the dominant mechanism for elevating $H_0$, while the energy transfer parameter $\gamma$ plays a complementary but secondary role, becoming effective only in the presence of non-zero $\Gamma(t)$.

For the dataset excluding SH0ES (Pantheon$+$ + CMB + BAO + GRB + CC), the analysis yields markedly different results compared to the SH0ES-inclusive dataset. When both $\gamma$ and $\Gamma_0$ are allowed to free, both Models I and II produce $H_0$ values lower than or consistent with $\Lambda$CDM, with Model I yielding $h = 0.683 \pm 0.013$ and Model II giving $h = 0.665 \pm 0.013$, compared to $h = 0.6838 \pm 0.0028$ for $\Lambda$CDM. The constrained values show Model I with $\Gamma_0 = -0.028^{+0.029}_{-0.010}$ and $\gamma = 0.0095^{+0.0034}_{-0.0094}$, while Model II exhibits $\Gamma_0 = -0.0038^{+0.014}_{-0.0080}$ and $\gamma = 0.0033^{+0.0012}_{-0.0033}$. When $\Gamma(t) = 0$ (Table \ref{Table_4}), the models continue to yield $h$ values below $\Lambda$CDM, with $\gamma$ values reduced to $\gamma = 0.00058^{+0.00023}_{-0.00053}$ for Model I and $\gamma = 0.00279^{+0.00076}_{-0.0023}$ for Model II. 

Examining the $\gamma = 0$ scenarios without SH0ES reveals a contrasting behavior to the SH0ES-inclusive case. When $\gamma = 0$ with $\Gamma_0 < 0$ (Table \ref{Table_7}), both models yield $H_0$ values slightly higher than $\Lambda$CDM: $h = 0.6927^{+0.0048}_{-0.0091}$ for Model I and $h = 0.6894^{+0.0038}_{-0.0065}$ for Model II, compared to $h = 0.6838 \pm 0.0031$ for $\Lambda$CDM. However, this modest elevation is insufficient to address any tension, as these values remain within the low-$H_0$ regime preferred by CMB+BAO data. The magnitude of $\Gamma_0$ is substantially reduced compared to the SH0ES-inclusive dataset: $\Gamma_0 = -0.00134^{+0.0013}_{-0.00052}$ for Model I and $\Gamma_0 = -0.00261^{+0.0026}_{-0.00095}$ for Model II, representing reductions of approximately one order of magnitude. When $\gamma = 0$ with $\Gamma_0 > 0$ (Table \ref{Table_8}), Model I yields $h = 0.673^{+0.011}_{-0.0058}$ and Model II gives $h = 0.663^{+0.014}_{-0.0069}$, both below $\Lambda$CDM, with positive $\Gamma_0$ values of $0.00177^{+0.00070}_{-0.0016}$ and $0.0124^{+0.0033}_{-0.0090}$, respectively. This demonstrates that without the high-$H_0$ SH0ES prior, neither matter annihilation ($\Gamma_0 < 0$) nor matter creation ($\Gamma_0 > 0$) can elevate the Hubble constant above $\Lambda$CDM levels, indicating that the models' ability to address the Hubble tension is fundamentally dependent on the inclusion of local distance ladder measurements that favor higher $H_0$ values.

For both models, the parameter $\gamma$ governs the energy transfer from cosmic fluids to the effective entropic dark energy. The constrained values of $\gamma$ in the presence of both parameters ($\gamma = 0.0128^{+0.0050}_{-0.012}$ for Model I and $\gamma = 0.00241^{+0.00084}_{-0.0024}$ for Model II with Pantheon$+$$\&$SH0ES data) indicate a relatively slow energy transfer process, suggesting that the effective entropic dark energy evolves gradually. Crucially, the data consistently favor negative values of $\Gamma_0$ across both models when the Hubble tension is reduced: $\Gamma_0 = -0.043^{+0.037}_{-0.015}$ for Model I and $\Gamma_0 = -0.0211^{+0.011}_{-0.0049}$ for Model II. This preference for $\Gamma_0 < 0$ unambiguously indicates that matter annihilation, rather than creation, is essential for addressing the tension. The interplay between matter annihilation (quantified by $\Gamma_0$) and energy transfer (quantified by $\gamma$) is evident from the strong degeneracy observed when either parameter is set to zero: when $\gamma = 0$ with $\Gamma_0 < 0$, the magnitude of $\Gamma_0$ reduces substantially (by factors of approximately 9 and 2 for Models I and II, respectively), yet both models maintain their ability to alleviate the Hubble tension. Conversely, when $\Gamma(t) = 0$, even with non-zero $\gamma$ values ($\gamma = 0.000093^{+0.000031}_{-0.000095}$ for Model I and $\gamma = 0.000245^{+0.000084}_{-0.00025}$ for Model II), the models yield $H_0$ values consistent with $\Lambda$CDM, failing to address the tension. This hierarchy establishes that matter annihilation is the primary driver, while energy transfer plays a complementary role.

As articulated by \cite{Halder:2025eze}, the condition $\Gamma(t)<0$ alone is insufficient to explain accelerated cosmic acceleration unless accompanied by a novel fluid with a negative equation of state. Our results corroborate this premise: the effective entropic dark energy, generated through energy transfer from cosmic fluids to the cosmological horizon, fulfills this role. In scenarios where $\gamma = 0$ with $\Gamma_0 < 0$, the entropic dark energy density $\rho_e$ effectively acts as a cosmological constant with equation of state $w_e = -1$, serving as an additional component that, in conjunction with matter annihilation, modifies late-time expansion. Conversely, when $\gamma = 0$ with $\Gamma_0 > 0$ (matter creation), both models fail to resolve the tension, yielding $H_0$ values ($h \approx 0.687$) indistinguishable from $\Lambda$CDM, with extremely small positive $\Gamma_0$ values that are orders of magnitude smaller than their negative counterparts. This demonstrates that the sign of $\Gamma_0$ is critical: only matter annihilation ($\Gamma_0 < 0$), combined with energy transfer to entropic dark energy, can effectively alleviate the Hubble tension.

Furthermore, our framework substantiates the proposition that solutions confined solely to either early or late phases are insufficient to resolve the Hubble tension, as delineated in \cite{Vagnozzi:2023nrq}. The combined effects of matter annihilation, energy transfer, and the resulting modification to both early universe dynamics (through altered $\Omega_m$ and $\Omega_b$ values) and late-time expansion (through enhanced effective dark energy) are necessary to achieve the observed mitigation. Notably, this resolution is strongly dependent on the inclusion of local distance ladder measurements: when SH0ES data are excluded (Pantheon$+$+CMB+BAO+GRB+CC), both models yield $H_0$ values ($h = 0.683 \pm 0.013$ for Model I and $h = 0.665 \pm 0.013$ for Model II) that are lower than or consistent with $\Lambda$CDM, demonstrating that the models' efficacy requires the high-$H_0$ constraint from local measurements.

The parameters $\gamma$ and $\Gamma_0$ exhibit a strong degeneracy mediated through their combined effect on the effective scaling parameter $\alpha = (1 - \gamma)(1 - \Gamma_0/3)$, which appears in the continuity equations governing cosmological evolution. For the Pantheon$+$$\&$SH0ES+CMB+BAO+GRB+CC dataset, we obtain $\alpha = 1.00131 \pm 0.00042$ for Model I and $\alpha = 1.0046 \pm 0.0014$ for Model II when both parameters are free. These small but significant deviations from unity modify early-universe physics, resulting in reduced sound horizons at both recombination and the baryon drag epoch compared to $\Lambda$CDM ($r_s(z_*) = 145.4 \pm 1.3$ Mpc, $r_s(z_d) = 148.0 \pm 1.4$ Mpc).

In Model I, the scaling parameter $\alpha = 1.00131$ modifies the baryon-to-photon ratio evolution to $R_b = \frac{3\Omega_{b,0}}{4\Omega_{\gamma,0}}(1+z)^{-\alpha}$ rather than the standard $(1+z)^{-1}$ scaling in $\Lambda$CDM. This leads to a slightly smaller $R_b$ at high redshifts, resulting in a marginally enhanced sound speed $c_s(z) = c/\sqrt{3(1+\overline{R}_b(1+z)^{-\alpha})}$ by approximately $0.9\%$ during the pre-recombination era. Simultaneously, the matter density scaling $\rho_m \propto a^{-3\alpha}$ and the early-time behavior of effective entropic dark energy $\rho_e$ (which mimics radiation and matter before recombination) enhance the Hubble function $H(z)$ by approximately $4$--$5\%$ at recombination. The integrated sound horizon $r_s(z_*) = \int_{z_*}^{\infty} c_s(z')/H(z') \, dz'$ is reduced because the enhancement to $H(z)$ in the denominator dominates over the modest increase in $c_s$ in the numerator. This yields $r_s(z_*) = 139.7 \pm 1.4$ Mpc and $r_s(z_d) = 142.1 \pm 1.4$ Mpc, representing reductions of $4.1\%$ and $4.0\%$ respectively from $\Lambda$CDM. This mechanism allows the CMB angular acoustic scale $\theta_* = r_s(z_*)/r(z_*)$ to remain consistent with Planck observations (through the shift parameters $R$ and $l_a$) while permitting higher $H_0$ values.

In Model II, where only dark matter undergoes creation/annihilation while baryons and radiation transfer energy to dark matter and effective entropic dark energy, the baryon-to-photon ratio retains approximately standard $(1+z)^{-1}$ scaling since baryons are not directly affected by $\Gamma_0$. Consequently, the sound speed remains close to $\Lambda$CDM values. The sound horizon reduction emerges entirely through modifications to the Hubble function arising from dark matter dynamics, energy transfer mechanisms, and the early-time contribution of $\rho_e$. The smaller scaling parameter deviation ($\alpha = 1.0046$ versus $1.00131$ in Model I) leads to a more modest $H(z)$ enhancement of approximately $2.5$--$3\%$, yielding $r_s(z_*) = 141.5 \pm 1.4$ Mpc and $r_s(z_d) = 143.8 \pm 1.4$ Mpc (reductions of $2.7\%$ and $2.8\%$ from $\Lambda$CDM).

To elucidate the individual contributions of $\gamma$ and $\Gamma_0$, we examine scenarios where $\gamma = 0$ with $\Gamma_0 < 0$, isolating the matter annihilation mechanism. Remarkably, Model I yields $\alpha = 1.00152 \pm 0.00038$ with sound horizons $r_s(z_*) = 140.1 \pm 1.4$ Mpc and $r_s(z_d) = 142.5 \pm 1.4$ Mpc, differing by only $0.4$ Mpc from the $\gamma \neq 0$ case despite a ninefold reduction in $|\Gamma_0|$ (from $-0.043$ to $-0.0046$). This remarkable consistency demonstrates that matter annihilation alone is sufficient to achieve the necessary sound horizon reduction, with the larger $\alpha$ value compensating for the smaller $|\Gamma_0|$ magnitude. Model II exhibits similar robustness: with $\gamma = 0$, we obtain $\alpha = 1.0038 \pm 0.0012$ and sound horizons $r_s(z_*) = 141.4 \pm 1.3$ Mpc and $r_s(z_d) = 143.7 \pm 1.4$ Mpc, differing by only $0.1$ Mpc from the $\gamma \neq 0$ case. This indicates that dark matter annihilation, which releases energy feeding into effective entropic dark energy, sufficiently enhances $H(z)$ to maintain sound horizon reduction even without direct energy transfer from baryons and radiation.

The strong parameter degeneracy through $\alpha$ is evident: when $\gamma = 0$, the effective scaling becomes $\alpha = 1 - \Gamma_0/3$, and smaller magnitudes of $\Gamma_0$ can produce similar or even larger $\alpha$ values compared to cases with $\gamma \neq 0$. The data constrain the effective combination $\alpha$ more tightly than individual parameters, explaining why multiple paths in $(\gamma, \Gamma_0)$ parameter space yield nearly identical cosmological observables. Both models demonstrate that matter annihilation ($\Gamma_0 < 0$) is the primary mechanism for sound horizon reduction and Hubble tension mitigation, with energy transfer ($\gamma$) playing a complementary but not strictly essential role. Model I achieves larger sound horizon reductions through the dual mechanism of $R_b$ modification and $H(z)$ enhancement, while Model II relies solely on $H(z)$ modifications but remains effective. The consistency of results across different parameter configurations, with all sound horizon values lying within $1\sigma$ uncertainties of each other for a given model, confirms that $\alpha$ encodes the essential physics determining early-universe dynamics in these thermodynamic frameworks.

\begin{figure*}[t]
    \centering
    \includegraphics[width=\textwidth]{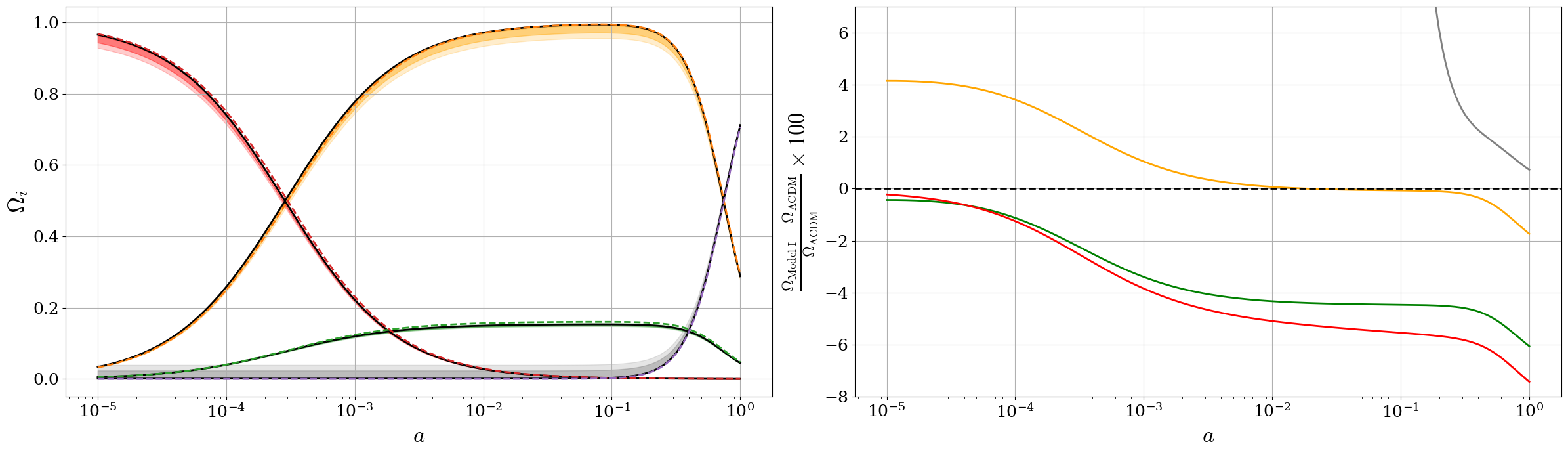}
    \caption{Evolution of the dimensionless density parameters $\Omega_i(a)$ (left) with $1\sigma$ and $2\sigma$ confidence regions shown as shaded areas, and percent deviation of $\Omega_i$ relative to $\Lambda$CDM (right), as a function of the scale factor $a$ for Model I. The dashed lines represent $\Lambda$CDM, and solid lines correspond to best-fit values for Model I. Colors: orange (total matter), green (baryonic matter), red (radiation), gray (effective entropic dark energy). Data: Pantheon$+$$\&$SH0ES+CMB+BAO+GRB+CC with priors $\gamma > 0$ and $\Gamma_0$ free.}
    \label{model1}
\end{figure*}

\begin{figure*}[t]
    \centering
    \includegraphics[width=\textwidth]{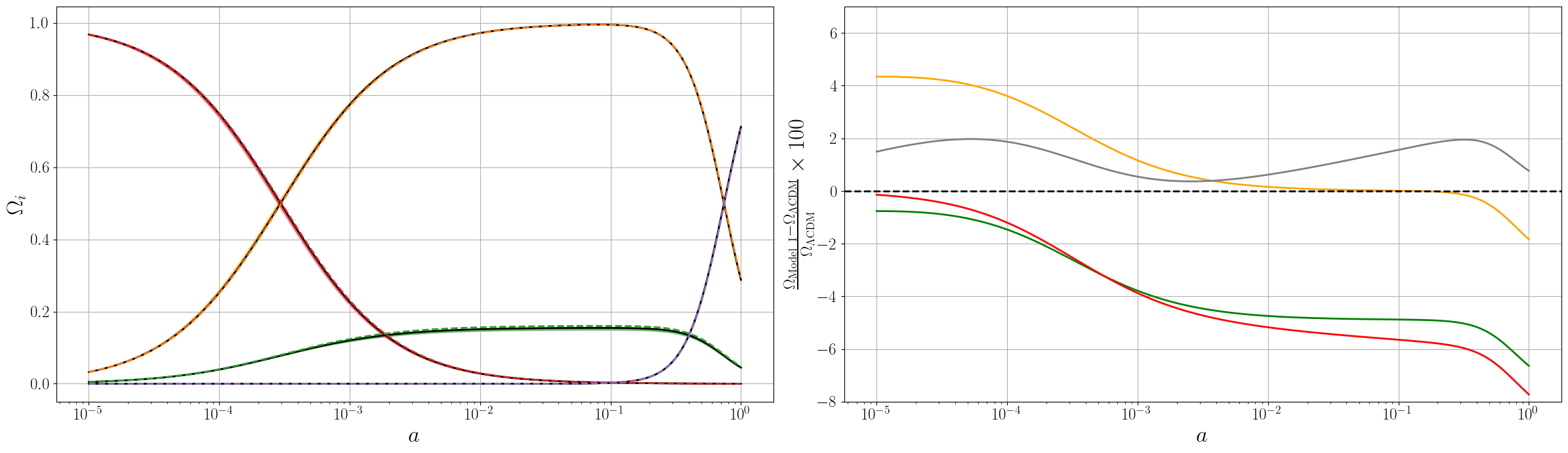}
    \caption{Same as Fig.~\ref{model1} but with priors $\gamma = 0$ and $\Gamma_0 < 0$.}
    \label{model1a}
\end{figure*}

\begin{figure*}[t]
    \centering
    \includegraphics[width=\textwidth]{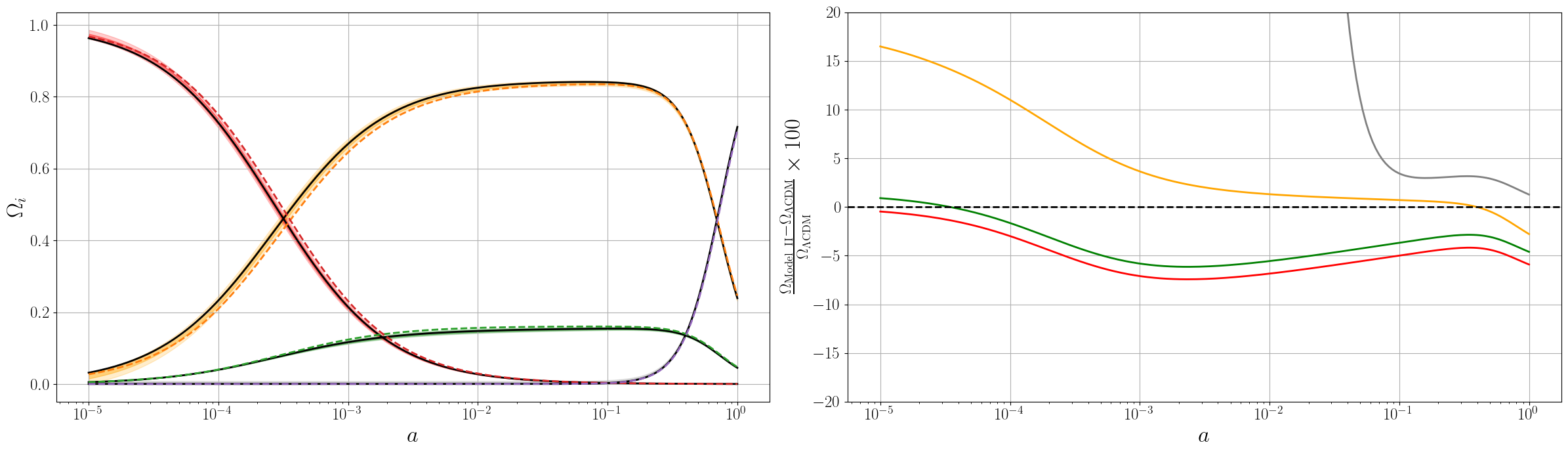}
    \caption{Evolution of the dimensionless density parameters $\Omega_i(a)$ (left) and percent deviation relative to $\Lambda$CDM (right) for Model II. Orange represents dark matter only (not total matter). Data: Pantheon$+$$\&$SH0ES+CMB+BAO+GRB+CC with priors $\gamma > 0$ and $\Gamma_0$ free. Other colors and conventions as in Fig.~\ref{model1}.}
    \label{model2}
\end{figure*}

\begin{figure*}[t]
    \centering
    \includegraphics[width=\textwidth]{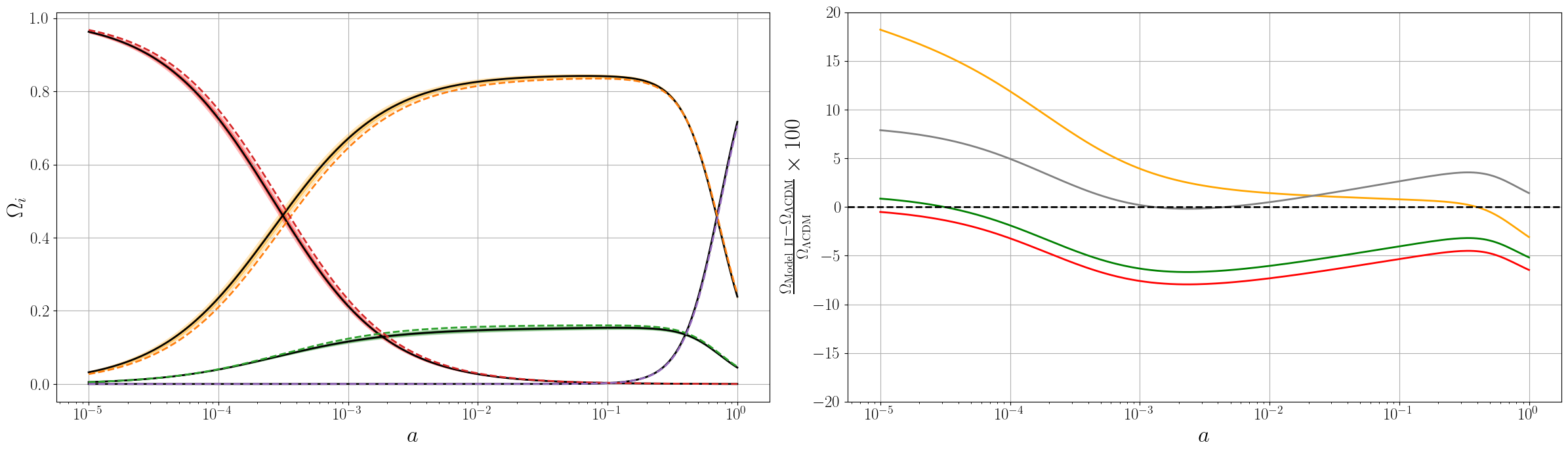}
    \caption{Same as Fig.~\ref{model2} but with priors $\gamma = 0$ and $\Gamma_0 < 0$. Orange represents dark matter only.}
    \label{model2a}
\end{figure*}
An analysis of the evolution of dimensionless density parameters for Model I (Figure \ref{model1}) with both $\gamma$ and $\Gamma_0$ free reveals that total matter exceeds $\Lambda$CDM before recombination, while baryonic matter and radiation are comparatively reduced. This divergence arises from matter creation/annihilation and energy transfer to effective entropic dark energy. Notably, $\rho_e$ exhibits very large positive deviations at early times, indicating substantial modification to the energy density composition. In the post-recombination era, continuous energy transfer from matter and radiation to $\rho_e$ becomes evident, with effective entropic dark energy remaining substantially elevated at late times. The pronounced positive deviation in $\rho_e$ during late times highlights the expansion rate modification that addresses the Hubble tension through elevated $H_0$ values.

Model II (Figure \ref{model2}) exhibits notably different behavior: baryonic matter and radiation initially exceed $\Lambda$CDM before displaying precipitous decline, indicating rapid energy transfer from baryons and radiation to dark matter, particularly pronounced around recombination. Dark matter shows sharp positive deviation around recombination, subsequently decreasing and crossing into negative deviation at late times, counterbalanced by substantial increase in effective dark energy. This energy flow pattern—first from baryons/radiation to dark matter, then from all components to $\rho_e$—reflects the two-stage interaction structure unique to Model II.

For the constrained scenario with $\gamma = 0$ and $\Gamma_0 < 0$, the effective entropic dark energy exhibits markedly different evolution. In Model I (Figure \ref{model1a}), $\rho_e$ maintains a modest $\sim 2\%$ positive deviation before recombination, decreases to near-zero around recombination, then rises to $\sim 2\%$ at late times. This contrasts sharply with the unconstrained case where $\rho_e$ shows very large early-time deviations. The suppressed behavior reflects that without continuous energy transfer ($\gamma = 0$), $\rho_e$ is fed exclusively by matter annihilation through $\Gamma_0 = -0.0046$. Despite this modest evolution, the sound horizon reduces to $r_s(z_*) = 140.1 \pm 1.4$ Mpc (differing by only $0.4$ Mpc from the unconstrained case), and $H_0 = 71.63 \pm 0.78$ km s$^{-1}$ Mpc$^{-1}$ remains nearly identical, demonstrating that matter annihilation alone is sufficient for Hubble tension mitigation.

Model II with $\gamma = 0$ (Figure \ref{model2a}) exhibits distinctly different behavior: $\rho_e$ shows substantial $\sim 7\%$ positive deviation before recombination, significantly larger than Model I's $\sim 2\%$. This enhanced deviation arises from the two-stage mechanism where dark matter serves as an intermediate energy reservoir. Dark matter exhibits $\sim 15\%$ positive deviation before recombination, indicating it receives energy and subsequently transfers excess energy to $\rho_e$ through annihilation ($\Gamma_0 = -0.0114$). The larger $|\Gamma_0|$ in Model II compared to Model I ($0.0114$ vs $0.0046$), combined with the enriched dark matter density, produces the enhanced $7\%$ deviation in $\rho_e$ at early times. Around recombination, $\rho_e$ approaches near-zero deviation, then rises to $\sim 2\%$ at late times, mirroring Model I's late-time pattern. Despite the different early evolution ($7\%$ vs $2\%$), Model II achieves similar sound horizon reduction ($r_s(z_*) = 141.4 \pm 1.3$ Mpc) and Hubble constant.

The key distinction between models in the $\gamma = 0$ scenario lies in early-universe dynamics: Model I distributes annihilation effects uniformly across all species with smaller $|\Gamma_0|$, producing uniform $\sim 2\%$ deviations; Model II uses dark matter as an energy buffer with larger $|\Gamma_0|$, amplifying early-time deviations to $7\%$ in $\rho_e$ and $15\%$ in dark matter. Nevertheless, both models converge to similar late-time behaviors ($\sim 2\%$ deviations) and achieve comparable observational outcomes, demonstrating that the effective scaling parameter $\alpha$ ultimately governs cosmological dynamics regardless of specific energy flow pathways.

The comparison between scenarios reveals the complementary roles of the two mechanisms. When $\gamma \neq 0$, continuous energy transfer sustains dramatically amplified $\rho_e$ deviations throughout cosmic evolution. When $\gamma = 0$, $\rho_e$ depends solely on matter annihilation, producing modest deviations yet achieving nearly identical sound horizons and $H_0$ values. This remarkable consistency demonstrates that $\alpha$, rather than absolute $\rho_e$ magnitudes, governs cosmological dynamics. The parameter degeneracy allows different $(\gamma, \Gamma_0)$ combinations to produce similar $\alpha$ values: when $\gamma = 0$, the larger $\alpha = 1.00152$ (Model I) compensates for reduced matter annihilation and absence of continuous transfer. This confirms matter annihilation as the primary mechanism, with energy transfer serving as an amplifying but not essential component for Hubble tension resolution.

The model comparison statistics provide additional insights into the relative performance of Models I, II, and $\Lambda$CDM across different scenarios and datasets. For the Pantheon$+$$\&$SH0ES+CMB+BAO+GRB+CC dataset with both $\gamma$ and $\Gamma_0$ as free parameters, the Bayesian evidence yields $\ln B_{\Lambda\text{CDM,I}} = -2.43$ and $\ln B_{\Lambda\text{CDM,II}} = -5.99$, indicating that both models are disfavored relative to $\Lambda$CDM according to the Jeffreys scale (weak to moderate evidence against). However, the AIC tells a different story: $\Delta AIC = 11.19$ for Model I and $\Delta AIC = 5.32$ for Model II suggest that both models provide substantially better fits than $\Lambda$CDM when accounting for model complexity, with Model I showing stronger support. DIC shows similar trends, with $\Delta DIC = 19.26$ for Model I and $\Delta DIC = 5.86$ for Model II. When $\gamma = 0$ with $\Gamma_0 < 0$, Model I achieves positive Bayes factor ($\ln B_{\Lambda\text{CDM,I}} = 1.65$), indicating weak preference over $\Lambda$CDM, while maintaining substantial AIC and DIC improvements ($\Delta AIC = 13.31$, $\Delta DIC = 13.34$). In contrast, when $\gamma = 0$ with $\Gamma_0 > 0$ or when $\Gamma(t) = 0$, both models show strong negative evidence ($\ln B < -6$) and negative $\Delta AIC$ and $\Delta DIC$ values, confirming their inability to outperform $\Lambda$CDM in these configurations. 

For the dataset excluding SH0ES (Pantheon$+$+CMB+BAO+GRB+CC), the model selection criteria exhibit consistently negative assessments across all scenarios. With both parameters set as free parameters, the models show substantial negative Bayes factors ($\ln B_{\Lambda\text{CDM,I}} = -9.97$, $\ln B_{\Lambda\text{CDM,II}} = -8.96$) alongside negative $\Delta AIC$ values ($\Delta AIC = -4.06$ for Model I, $\Delta AIC = -2.15$ for Model II), indicating that without local distance ladder constraints, the additional model complexity is not justified. When $\gamma = 0$ with $\Gamma_0 < 0$ (Table \ref{Table_7}), both models continue to show negative evidence ($\ln B_{\Lambda\text{CDM,I}} = -6.31$, $\ln B_{\Lambda\text{CDM,II}} = -5.99$) and negative $\Delta AIC$ ($\Delta AIC = -2.02$ for both models), despite the presence of matter annihilation. Even when $\gamma = 0$ with $\Gamma_0 > 0$ (Table \ref{Table_8}), where Model II shows the least negative Bayes factor ($\ln B_{\Lambda\text{CDM,II}} = -3.28$), the $\Delta AIC$ values remain negative or marginal ($\Delta AIC = -1.98$ for Model I, $\Delta AIC = -0.02$ for Model II), and the DIC shows Model II slightly preferred ($\Delta DIC = 1.43$) while Model I remains disfavored ($\Delta DIC = -1.21$). The uniformly negative model selection statistics without SH0ES data, across all parameter configurations, underscore a critical finding: the models' viability is intrinsically tied to the high-$H_0$ measurements from the local distance ladder, and they do not provide improved fits to datasets dominated by early-universe and low-redshift non-SH0ES probes alone.

The tension between Bayesian evidence (which penalizes additional parameters through the Occam factor) and AIC/DIC (which balance goodness-of-fit against complexity differently) reflects the trade-off between achieving better fits to high-$H_0$ SH0ES data and maintaining parsimony. Overall, Model I demonstrates more consistent performance across information criteria when $\Gamma_0 < 0$ with SH0ES data included, suggesting it provides a more robust framework for addressing the Hubble tension despite the Bayesian evidence penalty. However, the stark contrast between SH0ES-inclusive and SH0ES-exclusive results highlights that these models are specifically tailored to accommodate the discrepancy between local and early-universe $H_0$ measurements, rather than providing a universally preferred cosmological framework independent of dataset composition.
\begin{figure}[htp]
    \includegraphics[width=\columnwidth]{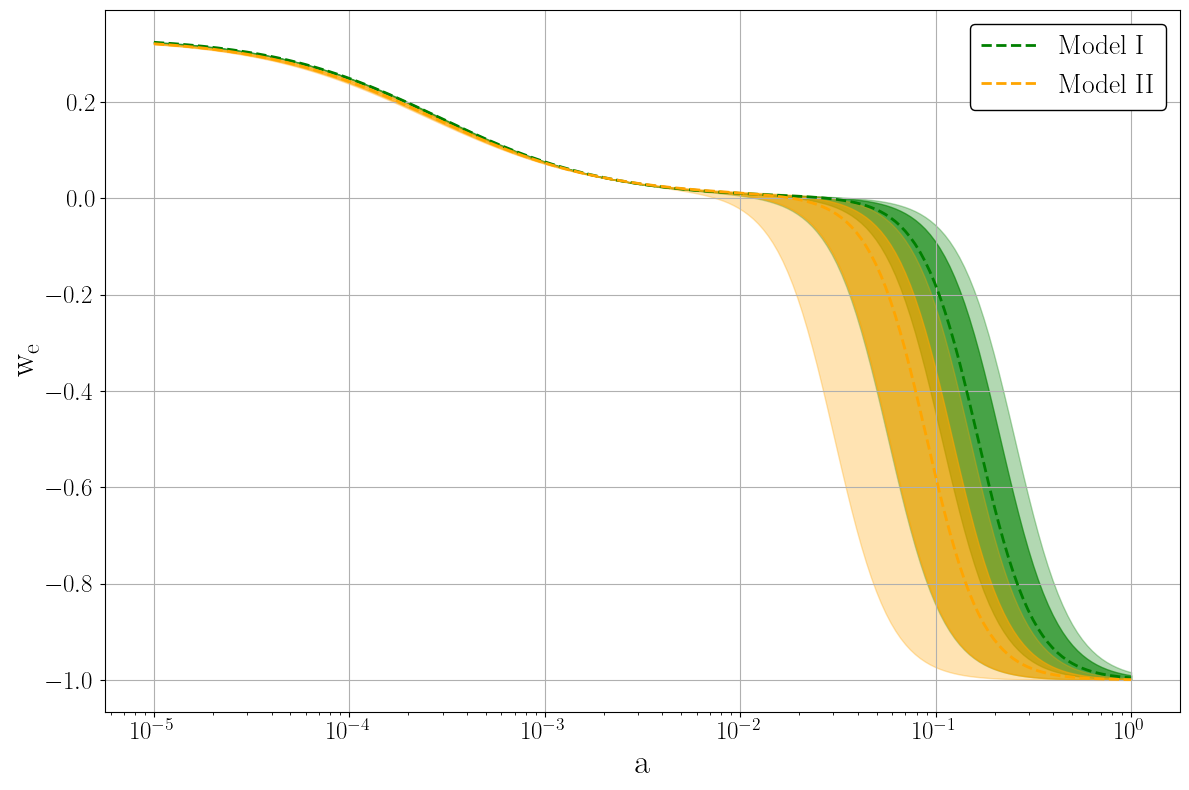}    
    \caption{Evolution of effective equation of state $w_e$ with respect to scale factor $a$ for Model I (green) and Model II (orange). Shaded regions are $1\sigma$ and $2\sigma$ confidence regions and doted lines represents the evolution on mean values. This plot applies to priors $\gamma > 0$, a free $\Gamma_0$, and Pantheon$+$\&SH0ES+CMB+BAO+GRB+CC dataset.   
}
\label{w_e1}
\end{figure}

An important aspect of these models is the behavior of the effective equation of state with respect to the effective entropic dark energy. As shown in Figure \ref{w_e1}, it starts exhibiting traits similar to radiation and transitions around the recombination era. Around $a=0.01$, it further evolves to display dust-like characteristics, afterwards behaving akin to quintessence, ultimately reaching the value of the cosmological constant today. In both models, the effective entropic dark energy undergoes changes depending on the different evolutionary stages of the universe, presenting a coherent scenario that extends from the matter and radiation sectors to the dark sector. This behavior is particularly evident in Chaplygin gas cosmological models \cite{Park:2009np, vonMarttens:2022xyr}, running vacuum models \cite{Moreno-Pulido:2022upl} and certain scalar-tensor gravity models \cite{Ferrari:2025egk}, where a single fluid initially mimics dark matter, eventually transitioning to represent dark energy. Thus, both dark matter and dark energy might be two manifestations of the same underlying entity. Specifically, in our framework, the effective entropic dark energy imitates not only early dark matter but also radiation before transitioning to the cosmological constant in the current era. In both models, this behavior is responsible for modifying the early physics around recombination, in addition to the energy flow among different constituents considered in these thermodynamic interacting scenarios.
Our Models I and II exhibit interesting connections with some of the models of running vacuum models (RVMs), which represent a well-established theoretical approach grounded in quantum field theory (QFT) in curved spacetime. The RVM framework is derived from rigorous renormalization-group calculations in QFT in curved spacetime, where the vacuum energy density $\rho_{\Lambda}$ and the gravitational coupling $G$ both exhibit logarithmic running with the energy scale $\mu \sim H$ \cite{Sola:2013gha, SolaPeracaula:2022hpd, Moreno-Pulido:2020anb, Moreno-Pulido:2022phq, SolaPeracaula:2023swx, MorenoPulido:2023jsc}. The canonical RVM describes the vacuum energy density as $\rho_{\Lambda}(H) = \rho_{\Lambda,0} + 3\nu M_{\rm Pl}^2 (H^2 - H_0^2)$, where the running parameter $\nu$ is constrained phenomenologically to the range $10^{-4}$ to $10^{-2}$ \cite{SolaPeracaula:2017esw, Sola:2016ecz, Sola:2017jbl, SolaPeracaula:2016qlq}. The RVM has been extensively tested against comprehensive cosmological data and has been shown to alleviate cosmological tensions, including both the Hubble and $\sigma_8$ tensions, often outperforming $\Lambda$CDM in information criteria such as $\Delta$AIC and $\Delta$BIC 
\cite{SolaPeracaula:2016jqh, SolaPeracaula:2017esw, Sola:2017jbl, Gomez-Valent:2017ywq, SolaPeracaula:2022hpd, Sola:2016ecz, SolaPeracaula:2016qlq}.

Under the specific conditions $\Gamma(t)=0$ (no matter creation/annihilation) and $m=1$ (Bekenstein entropy), our general framework reduces to equations that are formally identical to those of the canonical Type-I RVM, with our parameter $\gamma$ corresponding directly to the RVM parameter $\nu$. The bounds on $\gamma \approx 10^{-3}$--$10^{-2}$ derived from our analysis are consistent with the constraints on $\nu$ in RVM studies.\footnote{We note that in RVMs, $\nu$ may take negative values depending on the context, whereas in our framework, $\gamma$ is constrained to be non-negative by thermodynamic considerations related to entropy flow positivity.} 
Furthermore, the incorporation of particle production into the RVM framework \cite{SolaPeracaula:2019kfm} presents potential connections to our theoretical construct. However, a comprehensive and meticulous analysis is imperative for an accurate comparative assessment.

However, despite this mathematical equivalence in the specific case ($m=1$, $\Gamma(t)=0$), the two frameworks differ fundamentally in their theoretical motivations and scope. The RVM possesses rigorous quantum field theoretical foundations where the running of both $\rho_{\Lambda}$ and $G$ emerges automatically from renormalization-group analysis in curved spacetime—the running of $G$ is a built-in prediction of the QFT-based RVM, not an optional modification \cite{SolaPeracaula:2023swx, SolaPeracaula:2022hpd, MorenoPulido:2023jsc, Moreno-Pulido:2020anb, Moreno-Pulido:2022phq}. The RVM framework distinguishes between Type-I scenarios, where the vacuum energy density runs while $G$ is held approximately constant (to isolate effects), and Type-II scenarios (also called RRVM, renormalizable running vacuum model), where both $\rho_{\Lambda}$ and $G$ co-evolve, with matter densities following $\Lambda$CDM-like evolution while the vacuum-gravity sector carries the dynamical effects \cite{SolaPeracaula:2023swx}. Phenomenological analyses have shown that Type-II RVMs, where the running of $G$ plays a more prominent role, are particularly effective at alleviating the Hubble tension. 
Recent developments include the Stringy Running Vacuum Model, which incorporates gravitational anomalies and torsion from string theory, leading to RVM-type inflation and potential alleviation of both Hubble and structure growth tensions \cite{Mavromatos:2024pho, Mavromatos:2021bkw, Basilakos:2020qmu, Mavromatos:2021cdf}, and composite dark energy extensions where dark energy is a mixed fluid composed of running vacuum and additional components \cite{Gomez-Valent:2024tdb}.

In contrast, our approach is inspired by the holographic principle and employs thermodynamic reasoning to relate the bulk with matter creation/annihilation and horizon through the generalized entropy $S_m$ and the generalized mass-horizon relation (Eqs.~(\ref{S_m})--(\ref{ML})). The latter is not merely a heuristic ansatz but serves crucial purposes: it ensures thermodynamic consistency when applying generalized entropies with Hawking temperature (addressing long-standing inconsistencies in the literature \cite{Gohar:2023hnb,Cimdiker:2022ics,Gohar:2025yfx}), reduces to the Misner-Sharp mass for $\gamma=1/2$ and $m=1$ (connecting to well-established geometric quantities), and provides a systematic framework for exploring non-Bekenstein entropies. The parameter $\gamma$ in our framework represents an effective measure of energy flow across the horizon, constrained to be non-negative by thermodynamic considerations, rather than a quantum renormalization-group running parameter derived from QFT.

Our general setup aims to unify energy-flow and matter-creation/annihilation cosmologies using the first law of thermodynamics applied to open systems, as elaborated in Sections~II and~III. This provides a versatile structure that accommodates different models depending on the functional form of $\Gamma(t)$ and the parameter $m$. Importantly, for $m \neq 1$ (corresponding to Tsallis-Cirto entropy, Barrow entropy, or other generalized entropies) or for general matter creation/annihilation functions $\Gamma(t)$ beyond $\Gamma_0 H$, our framework does \textit{not} reduce to RVM—these represent genuine extensions exploring how different horizon entropy scalings and matter dynamics affect cosmology. The case we analyze in detail ($m=1$, $\Gamma(t)=\Gamma_0 H$) is a specific implementation within this broader framework.

A key empirical finding emerges when matter creation/annihilation is included. In our formulation with $m=1$ and $\Gamma(t)=\Gamma_0 H$, the data strongly favor negative $\Gamma_0$ values (matter annihilation) over positive values (matter creation) or $\Gamma_0=0$ (pure energy flow). As noted in two-fluid matter creation cosmologies~\cite{Halder:2025eze}, negative $\Gamma(t)$ typically requires an additional negative-pressure component to drive acceleration. In our framework, the energy transfer to effective entropic dark energy fulfills this role. Our analysis demonstrates that scenarios with $\Gamma(t)=0$ (pure energy-flow cosmologies, corresponding to some of the models of the RVMs with $\nu \sim \gamma$) are insufficient on their own to mitigate the Hubble tension when SH0ES data are excluded, while the combination of matter annihilation and energy transfer provides the necessary mechanism when local distance ladder measurements are included. This hierarchy—matter annihilation as primary, energy flow as complementary—parallels the role of running $G$ in Type-II RVMs, where modifications to the gravitational sector are essential for alleviating the Hubble tension.
From this perspective, the RVM represents a theoretically well-motivated and historically established framework with strong quantum field theoretical underpinnings and demonstrated success in fitting cosmological data \cite{SolaPeracaula:2016jqh, SolaPeracaula:2017esw, Sola:2017jbl, Gomez-Valent:2017ywq,  Sola:2016ecz, SolaPeracaula:2016qlq}. Our thermodynamic approach, by combining energy flow and matter creation/annihilation in a unified setting based on holographic principles, offers a complementary avenue that jointly modifies early- and late-time dynamics. Both frameworks share the common feature of time-varying dark energy and can address cosmological tensions, but they approach the problem from different fundamental principles—quantum field theory in curved spacetime versus thermodynamic holography applied to cosmological horizons with matter creation/annihilation. The fact that both converge mathematically for specific parameter choices ($m=1$, $\Gamma(t)=0$) while diverging in their theoretical foundations and broader scope demonstrates that multiple theoretical pathways may lead to observationally viable modifications of $\Lambda$CDM. This is consistent with arguments in~\cite{Vagnozzi:2023nrq} that resolving cosmological tensions may require simultaneous changes to both early and late cosmic epochs.
Furthermore, the cosmological inferences obtained in these scenarios depend sensitively on how BAO information is incorporated. In practice, BAO measurements are often compressed either into isotropic (angle-averaged) constraints on a single distance combination such as $D_V/r_d$, or into anisotropic constraints that separately probe the transverse and radial scales. While both approaches rely on a fiducial cosmology and are derived from the same underlying galaxy surveys, they differ in constraining power and parameter degeneracies.
The isotropic compression typically yields broader constraints, allowing greater freedom for extended models and consequently a wider range of inferred $H_0$ values when combined with late-time probes~ \cite{Gomez-Valent:2024tdb, Gomez-Valent:2024ejh}. By contrast, anisotropic BAO analyses exploit additional geometrical information and generally provide tighter constraints, strengthening the inverse distance ladder and reducing the allowed parameter space. As a result, conclusions regarding the ability of dynamical or composite dark energy models~\cite{Gomez-Valent:2024ejh, deCruzPerez:2025dni} to alleviate the Hubble tension can depend non-trivially on the adopted BAO treatment. Clarifying the impact of these methodological choices and ensuring consistency between different BAO compressions remain important for robust assessments of physics beyond $\Lambda$CDM.

A critical consideration in cosmological models involving matter creation or annihilation is consistency with thermodynamic principles. The seminal work by Prigogine and collaborators \cite{Prigogine:1989zz, Prigogine1988} established that, within the framework of irreversible thermodynamics applied to open systems, the second law requires spacetime to transform into matter (matter creation, $\Gamma_0 > 0$), while the inverse transformation (matter annihilation, $\Gamma_0 < 0$) appears forbidden. However, our results demonstrate that $\Gamma_0 < 0$ is essential for alleviating the Hubble tension, seemingly contradicting this theorem. This apparent inconsistency can be reconciled through several key mechanisms present in our models. First, the inclusion of energy transfer (quantified by $\gamma$) from cosmic fluids to the effective entropic dark energy fundamentally alters the thermodynamic landscape. Recent studies \cite{Jamil2010, Cardenas:2018nem} have demonstrated that the generalized second law of thermodynamics (GSL)—which requires that the total entropy (matter plus horizon entropy) does not decrease—is satisfied in interacting dark sector models where energy flows between components, independently of the specific interaction form or equation-of-state parameters. Second, thermodynamic analyses \cite{Moradpour:2014hta} show that consistency with thermodynamic equilibrium necessitates the existence of a component with negative equation of state ($w < -1/3$), a role fulfilled by our effective entropic dark energy with $w_e \approx -1$. Third, and most importantly, matter annihilation can be thermodynamically viable under specific conditions \cite{Yu:2022qjs}: when coupled with a component possessing negative pressure (such as our entropic dark energy), the entropy production from the energy transfer mechanism and the horizon entropy can compensate for the entropy reduction associated with particle annihilation, ensuring $\dot{S}_{\text{total}} \geq 0$. In our framework, the coupled system of matter annihilation ($\Gamma_0 < 0$) and energy transfer ($\gamma > 0$) to entropic dark energy creates a thermodynamically consistent scenario where the effective dark energy component acts as an entropy reservoir. The fact that $\Gamma(t) = 0$ (Table \ref{Table_3}) or $\Gamma_0 > 0$ (Table \ref{Table_6}) fail to alleviate the Hubble tension while yielding parameters consistent with $\Lambda$CDM further supports this interpretation: only the specific combination of matter annihilation with energy transfer to a negative-pressure component satisfies both observational constraints and thermodynamic requirements. Thus, while isolated matter annihilation would indeed violate the second law, the interacting nature of our models—where annihilated matter's energy is transferred to the entropic dark energy—ensures thermodynamic consistency through net positive entropy production in the coupled system. A comprehensive thermodynamic analysis, including explicit calculation of the entropy production rates and verification that the generalized second law is satisfied throughout cosmic evolution for our specific interaction forms, would strengthen this argument and is a worthwhile direction for future work.

\section{Conclusion} 
We have developed a thermodynamic framework to address the Hubble tension by incorporating gravitationally induced matter creation/annihilation with energy transfer from the cosmic bulk to the cosmological horizon. This approach integrates matter creation/annihilation cosmologies with energy flow cosmologies into a unified framework, modifying both early-universe physics around recombination and late-time cosmic acceleration. We analyzed two distinct scenarios: Model I, which considers matter creation/annihilation across all species with energy transfer to effective entropic dark energy, and Model II, which focuses on dark matter creation/annihilation with energy transfer from baryonic matter and radiation to both dark matter and effective entropic dark energy.

Our analysis reveals a clear hierarchy in the mechanisms responsible for elevating $H_0$ when local distance ladder measurements are included: matter annihilation emerges as the primary driver, while energy transfer plays an essential complementary role. The strong parameter degeneracy between $\Gamma_0$ and $\gamma$ demonstrates that when energy transfer is absent, less matter annihilation is required to achieve similar cosmological effects, yet matter annihilation alone remains crucial when SH0ES data are incorporated. Conversely, pure energy flow without matter creation/annihilation proves insufficient. Critically, matter creation scenarios fail entirely, yielding results indistinguishable from $\Lambda$CDM, establishing that the sign of $\Gamma_0$ is fundamental to the models' behavior when high-$H_0$ measurements are present.

A key finding concerns the role of observational datasets: when local distance ladder measurements (SH0ES) are excluded, both models consistently yield Hubble constant values comparable to or below $\Lambda$CDM across all parameter configurations. This demonstrates that the models' ability to accommodate higher $H_0$ values is critically dependent on the inclusion of local distance ladder data. Without SH0ES measurements, there is no evidence from information criteria that these models provide improvement over $\Lambda$CDM, regardless of the value of $\Gamma_0$. Model comparison statistics reflect this complexity: while Bayesian evidence penalizes the additional parameters, AIC and DIC criteria favor the models only when matter annihilation is present and SH0ES data are included, illustrating the trade-off between improved fits to high-$H_0$ measurements and model parsimony.

The effective entropic dark energy in our framework exhibits rich dynamical behavior, mimicking radiation in the early universe, transitioning through a dust-like phase around recombination, subsequently behaving as quintessence, and ultimately approaching the cosmological constant value today. This evolution naturally implements the principle that modifications to multiple cosmic epochs may be relevant for the Hubble tension discussion.

An important theoretical consideration involves the thermodynamic consistency of matter annihilation, which appears to contradict Prigogine's theorem forbidding such processes under the second law of thermodynamics. However, this apparent conflict is resolved in our framework: the coupled system of matter annihilation with energy transfer to a negative-pressure component ensures thermodynamic viability through net positive entropy production. The generalized second law remains satisfied in interacting dark sector models where energy flows between components, and the presence of effective entropic dark energy with negative equation of state provides the necessary thermodynamic sink. Nevertheless, a comprehensive analysis including explicit entropy production rate calculations would strengthen this argument and remains a priority for future investigation. The scenario with $\Gamma_0 < 0$ (matter annihilation) is thermodynamically viable within this coupled framework, though understanding its microscopic origin remains an open theoretical question. This configuration accommodates higher $H_0$ values when SH0ES data are included.

Our framework exhibits interesting connections with running vacuum models, which possess well-established theoretical foundations in quantum field theory. Under specific conditions, our equations become formally identical to canonical RVMs, with consistent parameter constraints emerging from both approaches. However, the frameworks differ fundamentally in their theoretical origins: RVMs arise from renormalization-group effects in quantum field theory, while our approach is inspired by holographic principles and thermodynamic reasoning. Both represent complementary pathways demonstrating that multiple theoretical foundations---quantum field theory versus thermodynamic considerations with matter creation/annihilation---may lead to observationally viable modifications of $\Lambda$CDM.

While this analysis employed the phenomenological creation function $\Gamma(t) = \Gamma_0 H$ with Bekenstein entropy, several important extensions remain for future work. First, exploring alternative functional forms for $\Gamma(t)$ and extending to nonstandard entropies through appropriate parameterization of $m$ would provide deeper insights, though careful attention to thermodynamic consistency with Hawking temperature is essential. Second, incorporating cosmological perturbations, comprehensive CMB power spectrum analysis, and assessment of structure growth is necessary to evaluate these models' relevance to the $\sigma_8$ tension and broader cosmological challenges. Third, rigorous verification of the generalized second law throughout cosmic history with explicit entropy calculations would provide quantitative thermodynamic constraints. Fourth, investigating whether these models can simultaneously address multiple cosmological tensions would test their broader applicability beyond the Hubble tension.

In conclusion, we have presented thermodynamically interacting cosmological models incorporating matter annihilation and energy transfer that modify cosmic expansion history through both reversible and irreversible thermodynamic processes. Crucially, the models' capacity to accommodate Hubble constant values consistent with local distance ladder measurements is critically dependent on the inclusion of SH0ES data in the analysis. In the absence of such local measurements, the models yield $H_0$ values comparable to $\Lambda$CDM and show no evidence of improvement according to information criteria. This strong dataset dependence indicates that these models specifically target the local-global $H_0$ discrepancy when local distance ladder priors are incorporated, rather than providing a universally preferred cosmological framework independent of dataset composition. The interplay between matter annihilation and energy transfer to effective entropic dark energy provides a thermodynamically consistent pathway  within this context, underscoring the significance of thermodynamic considerations in contemporary cosmology while acknowledging that comprehensive resolution of cosmological tensions and their interpretation requires further theoretical and observational developments.
\section*{Acknowledgments}
\noindent
The author thanks Vincenzo Salzano for his valuable discussions on the data analysis.

\bibliographystyle{apsrev4-1}
\bibliography{ref}

\end{document}